\PassOptionsToPackage{dvipsnames}{xcolor}
\documentclass{template/bytedance}

\usepackage[utf8]{inputenc}
\usepackage{amsmath}
\usepackage{amsfonts}
\usepackage{float}

\usepackage{tikz}
\usetikzlibrary{shapes, shapes.geometric, arrows.meta, positioning, fit, backgrounds, calc}

\fancypagestyle{firststyle}{%
  \fancyhf{}%
  \fancyhead[L]{\vskip 4mm\includegraphics[width=43.3mm,height=10mm]{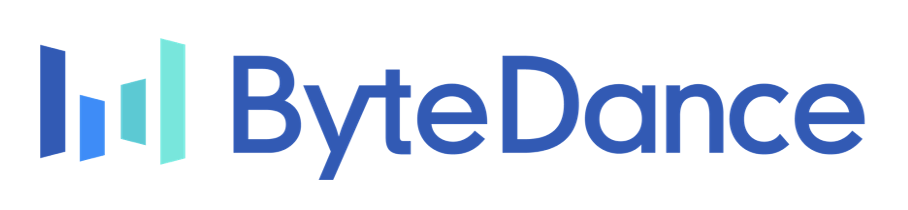}}%
}

\newcommand{\sysname}{\textsc{Tahoe}}

\usepackage{pifont}

\usepackage{listings}
\newif\ifcomments
\commentsfalse         

\newcommand{\diff}{\texttt{diff}}
\newcommand{\diffs}{\texttt{diffs}}

\title{TAHOE: Text-to-SQL with Automated Hint Optimization from Experience}

\author[1,2,*]{Zhiyi Chen}
\author[1]{Jie Song}
\author[1]{Peng Li}

\affiliation[1]{ByteDance Inc.}
\affiliation[2]{Georgia Institute of Technology}

\contribution[*]{Work done during an internship at the ByteBrain team, ByteDance}

\abstract{Large Language Models (LLMs) have democratized access to databases via Text-to-SQL, yet bridging the gap between prototypes and production remains challenging. Real-world deployments face strict dialect constraints (e.g., Snowflake), massive schemas, and evolving user preferences that vanilla LLMs struggle to handle. Existing solutions like Supervised Fine-Tuning (SFT) are costly and rigid, while manual prompt engineering and agentic test-time scaling are unscalable in latency and computation.

We present \sysname{}, a system that transforms prompt optimization into a dynamic data management problem. \sysname{} runs an error-driven hint learning pipeline over a two-phase lifecycle (Development and Deployment) to consolidate debugging traces into a structured Hint Bank. Compiler feedback is distilled into reusable \emph{Syntax Hints} that enforce dialect-specific rules, while execution and user feedback are converted into \emph{Semantic Hints} that capture schema- and user-specific logic. A novel \emph{Strategy Layer} models conflicting user intents as competing strategies under shared natural-language triggers; each strategy is annotated with a learning-time recency signal and, after learning, with post-learning attribution statistics that summarize its empirical success, harm, inertness, and support on actual generations. At inference time, \sysname{} performs context-aware retrieval and guides the LLM through a two-stage process: \emph{Logic Planning} to select and combine strategies, followed by \emph{SQL Synthesis} to generate dialect-correct queries. We implement and evaluate the development-phase workflow, while leaving deployment-time human-feedback updates for future work.

Evaluations on the Spider~2.0--Snow benchmark show that \sysname{} substantially improves hint-guided Text-to-SQL without updating model parameters. In the development-phase evaluation on the 113 supervised Spider~2.0--Snow-0212 examples, with our primary backbone (GPT-5.5), \sysname{} lifts pass rate from 61.95\% to 79.42\% and pass@4 from 72.57\% to 87.61\%, drives the Snowflake syntax pass rate to 100\%, and reduces average compiler-feedback critic rounds from 2.79 to 0.12 per sampled candidate. The same Hint Bank, plugged into weaker backbones without retraining, also yields large gains (e.g., $+19.7$~pp pass rate on Doubao-2.0-lite), demonstrating cross-model transferability. On held-out examples, syntax transfer remains strong, while semantic gains are more modest, suggesting that semantic benefits depend on how well the development set covers the target workload.
By converting transient feedback into a persistent and interpretable Hint Bank, \sysname{} provides a practical framework for robust and adaptable database interfaces.}

\date{\today}
\correspondence{Zhiyi Chen, \email{zchen798@gatech.edu}}

\begin{document}

\maketitle

\section{Introduction}

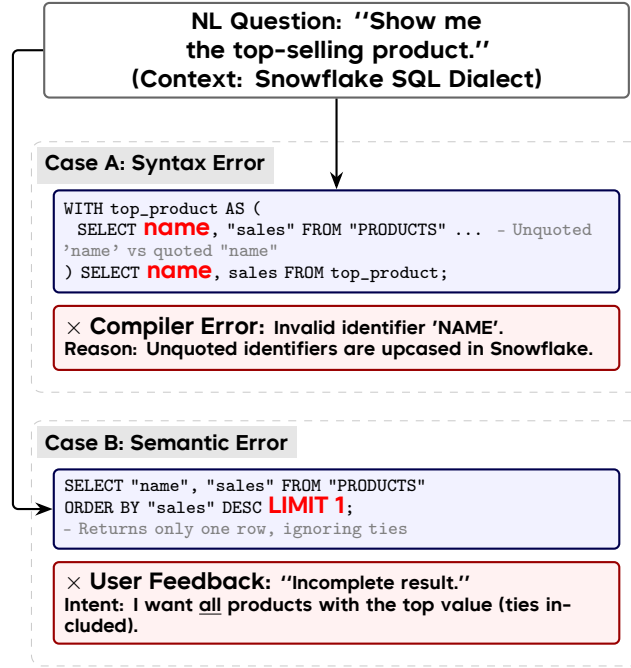
\begin{figure}[t]
\centering

\tikzset{
    container/.style={rectangle, draw=gray!40, dashed, rounded corners=4pt, inner sep=8pt, minimum width=7.5cm},
    titleBox/.style={rectangle, fill=gray!20, text=black, font=\bfseries\sffamily\footnotesize, inner sep=3pt, anchor=north west},
    sqlBox/.style={rectangle, draw=blue!30!black, fill=blue!5, thick, rounded corners=2pt, font=\ttfamily\scriptsize, align=left, inner sep=4pt, text width=7.2cm},
    errorBox/.style={rectangle, draw=red!60!black, fill=red!5, thick, rounded corners=2pt, font=\sffamily\scriptsize, align=left, inner sep=4pt, text width=7.2cm},
    arrow/.style={->, >=Stealth, thick, rounded corners=2pt}
}

\begin{tikzpicture}[node distance=0.3cm]

    \node[rectangle, draw=black!60, fill=white, rounded corners=2pt, thick, align=center, font=\sffamily\small, text width=7.5cm] (input) {
        \textbf{NL Question:} ``Show me the top-selling product.''\\
        \textit{(Context: Snowflake SQL Dialect)}
    };

    
    \node[sqlBox, below=1.2cm of input] (syn_sql) {
        WITH top\_product AS (\\
        \ \ SELECT \textcolor{red}{\textbf{name}}, "sales" FROM "PRODUCTS" \dots
        \ \ \textcolor{gray}{-- Unquoted 'name' vs quoted "name"}\\
        ) SELECT \textcolor{red}{\textbf{name}}, sales FROM top\_product;
    };
    
    \node[errorBox, below=0.15cm of syn_sql] (syn_err) {
        \textbf{$\times$ Compiler Error:} Invalid identifier 'NAME'.\\
        \textit{Reason: Unquoted identifiers are upcased in Snowflake.}
    };
    
    \begin{scope}[on background layer]
        \node (box_a_top) [above=0.1cm of syn_sql] {};
        \node[container, fit=(syn_sql) (syn_err) (box_a_top)] (box_a) {};
        
        \node[titleBox] at ([xshift=2pt, yshift=-2pt]box_a.north west) {Case A: Syntax Error};
    \end{scope}

    \draw[arrow] (input) -- (syn_sql);

    
    \node[sqlBox, below=1.0cm of box_a] (sem_sql) {
        SELECT "name", "sales" FROM "PRODUCTS"\\
        ORDER BY "sales" DESC \textcolor{red}{\textbf{LIMIT 1}};\\
        \textcolor{gray}{-- Returns only one row, ignoring ties}
    };
    
    \node[errorBox, below=0.15cm of sem_sql] (sem_fb) {
        \textbf{$\times$ User Feedback:} ``Incomplete result.''\\
        \textit{Intent: I want \underline{all} products with the top value (ties included).}
    };
    
    \begin{scope}[on background layer]
        \node (box_b_top) [above=0.1cm of sem_sql] {};
        \node[container, fit=(sem_sql) (sem_fb) (box_b_top)] (box_b) {};
        
        \node[titleBox] at ([xshift=2pt, yshift=-2pt]box_b.north west) {Case B: Semantic Error};
    \end{scope}

    \draw[arrow] (input.west) -- ++(-0.4,0) |- (sem_sql.west);

\end{tikzpicture}
\caption{Motivating Example. \textbf{Case A}: A syntax error where the model fails Snowflake's case-sensitivity rules. \textbf{Case B}: A semantic error where the model uses \texttt{LIMIT 1} instead of handling ties, mismatching user intent.}
\label{fig:motivation}
\end{figure}

Translating natural language (NL) questions into executable SQL queries, or \emph{Text-to-SQL}, is a long-standing goal of database research. Recent advances in large language models (LLMs) have brought remarkable progress. However, a significant gap persists between prototype demos and realistic deployments. In real-world scenarios---characterized by strict dialects (e.g., Snowflake), massive schemas, and evolving user preferences---generic LLMs frequently fail. As illustrated in Figure~\ref{fig:motivation}, models often violate dialect constraints (e.g., failing to quote mixed-case identifiers like \texttt{"name"} in Snowflake) or misinterpret subtle user intents (e.g., using \texttt{LIMIT 1} for ``top product'' instead of handling ties). These failures are not merely corner cases; on the recent Spider~2.0--Snow benchmark, even top-tier coding models (e.g., Qwen-Coder) achieve only $\approx$30\% execution accuracy, with general-purpose models like GPT-4o often performing worse due to a lack of domain-specific discipline~\cite{spider2025snow}.

To bridge this gap, recent research has diverged into several paradigms. As summarized in Table~\ref{tab:comparison}, while each offers partial solutions, they face distinct limitations in production environments:

\textbf{The ``Compute Trap'' (Test-Time Scaling):} A growing trend involves \emph{Test-Time Scaling} or agentic workflows, where models engage in multi-turn iterative refinement, generate massive numbers of candidates, or rely on ``fixer'' agents to correct errors~\cite{wang2025agentar}. While this can improve accuracy, it incurs substantial latency and financial costs---often requiring dozens of LLM calls for a single query. As highlighted in recent studies on long-term interaction, these systems suffer from \textbf{total amnesia}: they effectively ``reset'' between sessions, repeating the same errors on every new query and learning nothing from previous failures~\cite{chhikara2025mem0}. Consequently, an error fixed in one session is committed again in the next, necessitating redundant correction loops that waste computation and frustrate users.

\textbf{The ``Rigidity Trap'' (Supervised Fine-Tuning):} Another approach applies \emph{Supervised Fine-Tuning (SFT)} to internalize domain knowledge~\cite{scholak2021picard}. Although effective for static tasks, SFT presents significant data management challenges in dynamic database environments. Adapting to a schema change, a new SQL dialect, or evolving user preferences introduced by newly onboarded users requires expensive retraining. Furthermore, the learned weights are opaque and non-transferable; switching the base model renders the SFT investment obsolete.

\textbf{The ``Context Noise'' (Documentation RAG):} Finally, simple Retrieval-Augmented Generation (RAG) over database documentation or schemas often fails due to the ``Context Window Fallacy.'' As recent memory research indicates, simply extending context windows does not ensure effective utilization of information due to \textbf{attention degradation} over distant tokens~\cite{chhikara2025mem0, gurawa2025balancing}. Retrieving excessive raw schema descriptions introduces significant noise, leading to hallucinations when critical information is buried under voluminous metadata. These methods lack the precision to distill documentation into the actionable, concise logic required for complex reasoning.

In this paper, we propose \sysname{}, a system that addresses these challenges by transforming prompt optimization from a transient computational process into a persistent \textbf{data management} problem. Our core insight is that neither raw documentation retrieval (too noisy) nor iterative agentic reasoning (too slow and repetitive) is the right trade-off for production. Instead, we identify that retrieving \textbf{concise, error-driven hints}---summarized by LLM agents from past successful debugging---offers a balance of precision and efficiency. Unlike raw RAG that floods the context, \sysname{} retrieves only the distilled logic needed to solve the current problem. For instance, rather than retrieving pages of SQL dialect docs, the system might provide a specific syntax hint: \emph{Quote every database, schema, table, column, CTE, and alias exactly as it is stored; quote each element of a fully-qualified path separately.} 
Similarly, for ambiguous business logic, it retrieves precise semantic guidance: \emph{in a Google Analytics 4 (GA4) e-commerce schema, use \texttt{user\_pseudo\_id} instead of \texttt{user\_id} to uniquely identify customers} (see Section~\ref{sec:casestudy} for a concrete walkthrough).

\begin{table*}[t]
\centering
\caption{Comparison of \sysname{} with existing Text-to-SQL adaptation paradigms.}
\label{tab:comparison}
\resizebox{\textwidth}{!}{%
\begin{tabular}{lcccccc}
\toprule
\textbf{Paradigm} & \textbf{Representative Works} & \textbf{Latency} & \textbf{Continuous Evolution} & \textbf{Model Agnosticism} & \textbf{Generalizability} & \textbf{Accuracy} \\
& & \textit{(Efficiency)} & \textit{(Learns from Errors?)} & \textit{(Switchable Base?)} & \textit{(Avoids Overfitting?)} & \textit{(Performance)} \\ \midrule
\textbf{Vanilla LLMs} & Qwen-Coder~\cite{yang2025qwen3} & \textbf{Low} & Low & N/A & \textbf{High} & Low \\ \addlinespace
\textbf{Supervised Fine-Tuning (SFT)} & PICARD~\cite{scholak2021picard} & \textbf{Low} & Low & Low & Low & \textbf{High} \\ \addlinespace
\textbf{Test-Time Scaling (Agents)} & Agentar-Scale-SQL~\cite{wang2025agentar} & High & Low & \textbf{High} & \textbf{High} & \textbf{High} \\ \addlinespace
\textbf{Retrieval-Augmented (RAG)} & SQLGenie~\cite{ghosh2025sqlgenie} & \textbf{Low} & Medium & \textbf{High} & Medium & Medium \\ \midrule
\textbf{\sysname{} (Ours)} & -- & \textbf{Low} & \textbf{High} & \textbf{High} & \textbf{High} & \textbf{High} \\ \bottomrule
\end{tabular}%
}
\end{table*}

Crucially, \sysname{} addresses a major limitation of existing RAG and memory systems: the inability to handle \textbf{dynamic and conflicting interpretations}. Real-world queries often contain ambiguities where different users or contexts demand different valid solutions. Standard RAG treats these variations as noise or retrieves them randomly (the ``static retrieval'' problem). In contrast, we design a structured \textbf{Semantic Hint} model that maps a single NL Trigger to multiple competing \textbf{Strategies}. Each strategy explicitly encapsulates a distinct solution path with a rationale and contrastive examples. This allows \sysname{} to store conflicts explicitly and resolve them at runtime by ranking strategies based on flexible metrics, including a learning-time recency signal and post-learning attribution statistics that summarize each strategy's empirical success, harm, inertness, and support, ultimately offering the model a more reliable or user-preferred solution.

Beyond resolving ambiguity, constructing this structured \emph{Hint Bank} gives \sysname{} three advantages that are especially valuable for enterprise deployment: \textbf{Continuous Evolution}, \textbf{Model Agnosticism}, and \textbf{Generalizability}. First, because hints are external and human-readable, users can audit, revise, and extend them directly, allowing the system to continuously evolve as new errors, schemas, and user preferences emerge. Second, the Hint Bank is model-agnostic: unlike black-box SFT weights, it remains reusable even when the base LLM changes. Third, \sysname{} improves generalizability by decoupling \emph{Syntax Hints} from \emph{Semantic Hints} and organizing semantic logic by scope (\emph{General}, \emph{Database-specific}, and \emph{User-specific}), preventing localized conventions from contaminating unrelated settings. As a result, the system can adapt by updating or re-configuring the relevant hint subset rather than retraining model parameters.

To operationalize this idea, \sysname{} uses an \textbf{error-driven hint learning pipeline} to extract 
reusable syntax and semantic hints from failures and consolidate them into the Hint Bank. 
Hint learning is performed \emph{outside the critical inference path}: the Hint Bank is initialized 
before deployment (i.e., development phase) and further refreshed during deployment through periodic 
batch updates over accumulated user feedback. This design separates background hint construction and 
maintenance from online query serving. 
At query time, \sysname{} combines lightweight hint retrieval with two-stage hint-guided generation, 
reducing latency and monetary cost while mitigating repeated user-facing failures. 
\sysname{} is orthogonal to schema linking and can be seamlessly integrated with existing schema pruning techniques to support large databases. This paper focuses on validating the development-phase hint learning and reuse pipeline; deployment-time human-feedback refresh is part of the architecture but is not empirically evaluated in this benchmark setting.

Our contributions are summarized as follows: 

\begin{itemize} 
\item \textbf{Hint-Based Long-Term Memory Paradigm:} We introduce a hint-based long-term memory paradigm for Text-to-SQL that reframes 
prompt optimization as managing a persistent \emph{Hint Bank}. \sysname{} employs an error-driven hint learning pipeline that consolidates 
debugging traces and accumulated feedback into reusable Syntax and Semantic Hints, while keeping hint learning and maintenance outside 
the critical inference path. This design shifts most reasoning and correction overhead to background construction and periodic batch 
refresh, thereby mitigating context noise and alleviating the latency bottlenecks of agentic workflows while preserving compatibility 
with different base LLMs.

\item \textbf{Structured and Transferable Hint Architecture:} We design a modular Hint Bank that decouples syntax from semantics and 
organizes semantic logic into hierarchical scopes (General, Database-specific, User-specific). A trigger--strategy abstraction 
captures multiple competing solution paths under the same natural-language trigger and ranks them using flexible signals 
(e.g., recency and success rate). 
This structure makes the Hint Bank both \emph{interpretable} and \emph{model-agnostic}, and adaptable to diverse industry 
contexts without expensive retraining.

\item \textbf{Empirical Effectiveness:} On the 113 supervised development examples of Spider~2.0--Snow-0212, \sysname{} substantially 
outperforms strong vanilla baselines for our primary backbone GPT-5.5 (pass rate $61.95\%\!\to\!79.42\%$, pass@4 $72.57\%\!\to\!87.61\%$, 
syntax pass rate $96.24\%\!\to\!100\%$, avg.\ critic rounds $2.79\!\to\!0.12$ per sampled candidate), and the same Hint Bank also 
lifts weaker backbones (Doubao-2.0-lite, GPT-5) by double-digit pp without any model-specific adaptation, evidencing cross-model 
transferability.
\end{itemize}

\section{Related Work}
\label{sec:related_work}

\subsection{Automatic Prompt Optimization}
Recent work in automatic prompt optimization aims to reduce the reliance on manual
prompt design by using automated search or reflective mechanisms.
AutoHint~\cite{sun2023autohint} automatically generates and refines textual
``hints'' for a target task by prompting an LLM to propose, evaluate, and update
task-specific guidance.
Other methods optimize prompts via gradient-inspired updates and search over
discrete prompt candidates, such as the ``Automatic Prompt Optimization with
Gradient Descent and Beam Search'' framework~\cite{pryzant2023apo}.
More recently, reflective prompt evolution approaches such as
GEPA~\cite{agrawal2025gepa} use natural-language reflection over rollouts and
genetic-style mutation of prompts, and show that prompt-space optimization can
outperform reinforcement learning in weight space on several benchmarks.

These methods demonstrate the promise of automated prompt design, but typically
operate at the level of a \emph{single} (or small set of) global prompts tuned
for a fixed task.
They inherently struggle with the context-dependent conflicts common in enterprise data (e.g., different users needing different logic for the same query).
In contrast, \sysname{} reframes prompt optimization from a transient search process into a persistent data management problem. 
Rather than converging to one ``golden prompt,'' \sysname{} manages a structured \emph{Hint Bank}. 
By keying hints by triggers, scopes, and strategies, \sysname{} enables query-specific retrieval and explicit conflict management, allowing the system to adapt dynamically to distinct user needs.

\subsection{Prompting and Retrieval for Text-to-SQL}
Prompt-based improvements for Text-to-SQL have been explored in both research
and practical systems. DIN-SQL~\cite{pourreza2023din} decomposes Text-to-SQL
into modular sub-tasks (e.g., schema linking, SQL sketching, self-correction)
and uses in-context prompting of LLMs for each step. DAIL-SQL~\cite{gao2023text}
systematically evaluates prompt design and proposes an integrated strategy with
an emphasis on token efficiency.

A parallel line of work augments prompting with retrieval. ReFSQL~\cite{zhang2023refsql}
introduces a structure-enhanced retrieval framework that retrieves samples with
comparable specific knowledge to aid generation. Beyond research prototypes,
industrial systems such as SQLGenie~\cite{ghosh2025sqlgenie} adopt an
example-based RAG design, retrieving verified query--SQL pairs from an
\emph{Example Bank}.

These retrieval-augmented approaches primarily provide \emph{implicit} guidance
through raw examples. This design faces the ``Context Noise'' challenge:
retrieving large or weakly matched examples can introduce irrelevant tokens,
ambiguous patterns, and conflicting logic~\cite{gurawa2025balancing}. \sysname{}
takes an orthogonal approach by retrieving \emph{explicit, abstracted hints}
derived from error analysis rather than raw (question, SQL) pairs. First, instead of
implicitly encoding dialect quirks through examples, \sysname{} factors
dialect-specific syntax knowledge into minimal reusable rules (\emph{Syntax Hints}).
Second, rather than collapsing all behaviors into a flat retrieval space,
\sysname{} organizes semantic hints by explicit \emph{Scope} (\emph{General},
\emph{Database-specific}, and \emph{User-specific}). Third, when multiple valid
behaviors exist, \sysname{} uses a \emph{Strategy Layer} to model mutually
exclusive alternatives and rank them using explicit signals---namely a
learning-time recency timestamp augmented by post-learning attribution
statistics (success/harm/inert/support summaries that capture empirical evidence
of when a strategy helped, hurt, or was merely retrieved)---instead of relying solely on
similarity scores.

\subsection{Test-Time Scaling and Agentic Workflows}
A growing trend in Text-to-SQL focuses on \emph{Test-Time Scaling}, which enhances performance by increasing computational spend during inference.
Agentar-Scale-SQL~\cite{wang2025agentar} represents the state-of-the-art, proposing an orchestrated framework that combines RL-enhanced reasoning, iterative refinement, and parallel tournament selection.
While achieving high accuracy on benchmarks like BIRD, these methods fundamentally trade latency for performance.
As noted by the authors of Agentar-Scale-SQL, the reliance on multiple LLM calls results in \emph{``substantial computational overhead,''} making them \emph{``less suitable for real-time applications''}.
In typical benchmark-oriented test-time scaling workflows, corrections are not persisted across sessions, so the system may repeat similar errors on new queries, learning little from previous failures.

\sysname{} addresses this by ``shifting left'' the computational burden.
Instead of relying on heavy online computation to correct errors \emph{at runtime}, we perform reasoning during the \emph{Development Phase} or batch updates.
By consolidating error corrections into a persistent Hint Bank, \sysname{} substitutes expensive iterative re-reasoning with efficient memory retrieval.
This enables the system to avoid recurring errors without the latency penalties of agentic loops.

\subsection{LLM Memory and Knowledge Management}
There is a growing body of work on equipping LLM-based systems with explicit
long-term or external memory.
The Generative Agents architecture~\cite{park2023generative} extends an LLM
with a memory stream that stores natural-language records of an agent's
experiences.
MemGPT~\cite{packer2023memgpt} introduces an OS-inspired hierarchical memory
system to support long-running conversations.
ChatDB~\cite{hu2023chatdb} augments LLMs with databases as symbolic memory for complex multi-hop reasoning.
More recently, Mem0~\cite{chhikara2025mem0} proposes a scalable memory layer that dynamically extracts and consolidates user preferences into a structured store, explicitly addressing the ``reset problem'' by detecting and resolving conflicts in long-term interactions.

These works share the goal of separating parametric knowledge from an explicit,
inspectable memory store.
\sysname{} applies a similar philosophy to Text-to-SQL but introduces a specialized structure tailored for ambiguity management: the Strategy Layer.
Many generic memory systems are optimized toward maintaining a single consistent user profile, whereas \sysname{} organizes hints hierarchically to explicitly preserve competing interpretations.
This structure allows the system to manage the lifecycle of knowledge effectively---preserving mutually exclusive behaviors (e.g., varying metric definitions) as parallel strategies rather than treating them as noise.
By tracking lightweight per-strategy signals---a learning-time recency timestamp and post-learning attribution statistics (success/harm/inert/support summaries that capture empirical evidence of when a strategy helped, hurt, or was merely retrieved)---\sysname{} dynamically resolves these semantic conflicts at inference time based on the specific user context.

\subsection{Other Approaches to Robust Text-to-SQL}
A large body of work tackles Text-to-SQL via supervised
learning (SFT).
Classical parsers like SyntaxSQLNet~\cite{yu2018syntaxsqlnet} and RAT-SQL~\cite{wang2020rat} train on labeled pairs to achieve strong performance.
Recent approaches like PICARD~\cite{scholak2021picard} fine-tune models with constrained decoding.
While effective for static tasks, SFT presents the ``Rigidity Trap'': adapting to a schema change or new dialect necessitates expensive retraining.
Furthermore, the learned weights are opaque and non-transferable.

By contrast, \sysname{} keeps the underlying LLM fixed and updates only the external Hint Bank.
This yields a lightweight adaptation strategy that is inherently model-agnostic.
Unlike SFT parameters, which are discarded when upgrading the base LLM, our Hint Bank persists as a reusable, transferable asset.
By decoupling syntax from semantics, \sysname{} ensures that semantic hints remain valid even if the SQL dialect changes, providing a level of modularity that parameter-based methods cannot match.

\section{Background and Problem Formulation}

\subsection{Text-to-SQL and Real-World Challenges}

The Text-to-SQL task aims to translate a natural language (NL) question $q$ and a database schema $S$ into an executable SQL query $y$. 
While vanilla LLMs perform well on academic benchmarks with standard SQL dialects and simplified schemas, their effectiveness degrades significantly in real-world deployments. 
As captured by the recent \emph{Spider~2.0--Snow} benchmark~\cite{spider2025snow}, realistic scenarios expose the intrinsic limitations of generic models---specifically their lack of domain-specific parametric knowledge and tendency to hallucinate---through three distinct challenges:

\begin{itemize}
    \item \textbf{Strict Dialect Adherence:} Industrial environments strictly enforce dialect rules that diverge from standard SQL. For instance, in Snowflake SQL, unquoted identifiers are automatically uppercased. A generic LLM, biased by its pre-training on standard SQL, lacks this specific dialect knowledge and often hallucinates valid ANSI-SQL syntax (e.g., unquoted names) that triggers immediate compilation errors in strict environments.
    
    \item \textbf{Massive Schema Reasoning:} Real-world databases contain hundreds of inter-related tables with obscure column names. The massive context required to describe such schemas often exceeds the model's effective attention span, leading to schema hallucinations where the model invents non-existent join paths or misinterprets column definitions.
    
    \item \textbf{Ambiguous and Evolving User Intent:} Unlike academic datasets where questions are precise, real users often use domain jargon or imply constraints (e.g., ``top products'' might imply distinct counting). Furthermore, these preferences are dynamic---definitions of metrics often evolve over time, creating a ``moving target'' that static model weights struggle to track without expensive retraining.
\end{itemize}

\subsection{Error Taxonomy}
To address these challenges systematically, we categorize Text-to-SQL failures into two orthogonal types based on their feedback source.
This taxonomy underpins the error-driven hint learning pipeline developed in Section~\ref{sec:learning_process}.

\begin{itemize}
    \item \textbf{Syntax Errors (Compiler-Level):} The generated SQL is invalid and rejected by the database engine. These errors stem from dialect mismatches (e.g., keywords, quoting, function names) rather than logic. They are objectively verifiable via \emph{compiler feedback}.
    \item \textbf{Semantic Errors (Execution-Level):} The SQL executes successfully but yields incorrect data. These errors arise from misaligned business logic or user intent (e.g., incorrect filters, wrong aggregation scope). Detecting them requires \emph{execution feedback} against ground truth or human verification.
\end{itemize}

Figure~\ref{fig:motivation} grounds this taxonomy: the \emph{top panel} is a syntax error (an unquoted Snowflake identifier rejected by the compiler), and the \emph{bottom panel} is a semantic error (the SQL compiles but uses \texttt{LIMIT 1}, mismatching the user's intent to include ties).

\subsection{Problem Formulation}
Traditional approaches treat Text-to-SQL as learning a direct mapping $f_\theta(q, S) \to y$, where $\theta$ represents the model parameters. In \sysname{}, we freeze the LLM parameters $\theta$ and instead optimize an external knowledge structure: the \textbf{Hint Bank} $\mathcal{H}$.

We formally define the inference process and the optimization objective as follows:

\noindent\textbf{Execution Equivalence ($\textsc{Exec}$):} Let $\textsc{Exec}(y, D)$ denote the result set obtained by executing SQL query $y$ on a database instance $D$ conforming to schema $S$. A generated query $\hat{y}$ is semantically correct with respect to the ground truth $y^*$ if and only if $\textsc{Exec}(\hat{y}, D) = \textsc{Exec}(y^*, D)$.

\noindent\textbf{Interaction Cost ($C$):} We define the interaction cost $C(\hat{y}, y^*)$ as the number of feedback rounds required to transform an initial prediction $\hat{y}$ into a correct query. 
In a vanilla setting, if $\hat{y}$ fails compilation, $C \ge 1$ (syntax correction rounds). If $\hat{y}$ executes but returns incorrect results, $C \ge 1$ (user feedback rounds). The ideal case is $C=0$ (correct on the first attempt).

\noindent\textbf{Optimization Objective:}
The Hint Bank decomposes into two disjoint registries, $\mathcal{H} = \mathcal{H}_{syn} \cup \mathcal{H}_{sem}$, that are accessed by different mechanisms at inference time. We therefore decompose the retrieval function as
\[
h(q, S, \mathcal{H}) = \mathcal{H}_{syn}^{(\text{dialect})} \cup \mathcal{R}_{sem}(q, S, \mathcal{H}_{sem}),
\]
where $\mathcal{H}_{syn}^{(\text{dialect})}$ is the complete syntax-hint registry for the target dialect (always injected, not retrieved), and $\mathcal{R}_{sem}$ performs scope filtering followed by trigger-based semantic retrieval over $\mathcal{H}_{sem}$. $\mathcal{R}_{sem}$ implies a conflict resolution mechanism: it does not merely fetch similar examples but ranks competing \emph{strategies} based on flexible signals---a learning-time recency timestamp and post-learning attribution summaries (success/harm/inert/support produced by a post-learning attribution pass)---to maximize alignment with current user needs (as detailed in Sections~\ref{sec:management} and~\ref{sec:inference}).
The LLM generation is modeled as a conditional probability distribution $P_\theta(y \mid q, S, h)$.

Our ultimate goal is to minimize the expected interaction cost $\mathbb{E}[C]$ during deployment.
Since the interaction cost is minimized (i.e., $C=0$) if and only if the initial generation is correct, minimizing the expected cost is equivalent to maximizing the expected accuracy of the hint-guided inference.
Therefore, we aim to maintain a Hint Bank $\mathcal{H}^*$ that maximizes:

\begin{equation}
    \mathcal{H}^* = \arg\max_{\mathcal{H}} \mathbb{E}_{(q, S, D, y^*) \sim \mathcal{P}} \left[ \mathbb{I}\left[ \textsc{Exec}(\hat{y}_0, D) = \textsc{Exec}(y^*, D) \right] \right]
\end{equation}
where $\hat{y}_0 \sim P_\theta(\cdot \mid q, S, h(q, S, \mathcal{H}))$ denotes a single sampled generation conditioned on the retrieved hints (before any compiler-feedback or execution-feedback loop). At evaluation time we separately report the candidate-level pass rate (averaged over multiple sampled $\hat{y}_0$) and the example-level pass@4 (any-of-$k$ correctness over $k=4$ samples).
By maximizing the probability that $\hat{y}_0$ is correct, we effectively obviate the need for expensive feedback loops (minimizing $\mathbb{E}[C]$ towards 0).

Crucially, the cost of optimizing $\mathcal{H}$ (e.g., error analysis, strategy merging) is incurred \emph{asynchronously}---during the Development Phase or via batch updates in deployment---so that it is decoupled from the synchronous user interaction.

\section{System Architecture}
\begin{figure}[t]
\centering
\resizebox{\textwidth}{!}{%
\begin{tikzpicture}[
    node distance=1.0cm and 2.0cm,
    font=\sffamily\small,
    >={Stealth[length=2mm]},
    process/.style={rectangle, draw=blue!60!black, fill=blue!5, thick, rounded corners=2pt, minimum height=0.9cm, minimum width=2.2cm, align=center},
    database/.style={cylinder, cylinder uses custom fill, shape border rotate=90, draw=black!60, fill=gray!10, aspect=0.25, minimum height=1.1cm, minimum width=1.3cm, align=center, font=\scriptsize},
    agent/.style={rectangle, draw=blue!40!black, fill=blue!10, thick, rounded corners=5pt, minimum height=0.9cm, minimum width=2.2cm, align=center}, 
    feedback/.style={dashed, draw=black, thick, ->, rounded corners=5pt},
    groupbox/.style={draw=gray!50, dashed, thick, rounded corners=12pt, fill=white},
    label/.style={font=\bfseries\sffamily\small, text=gray!80}
]


    \node[process, fill=yellow!10, draw=orange!40, thick, minimum height=3cm, minimum width=3.8cm] (hintbank) {
        \textbf{Hint Bank} ($\mathcal{H}$)\\
        \scriptsize
        \begin{tabular}{l}
        \\[-1ex]
        \textbf{1. Syntax Hints} ($\mathcal{H}_{syn}$) \\
        \quad \textit{(Dialect-specific)} \\[0.5ex]
        \textbf{2. Semantic Hints} ($\mathcal{H}_{sem}$) \\
        \quad - General / DB / User \\
        \quad - Strategy Layer \\
        \quad \textit{\hspace{2mm}(Conflict Resolution)}
        \end{tabular}
    };

    \node[process, left=1.2cm of hintbank] (manager) {Hint Management\\Module};
    \node[agent, left=0.8cm of manager] (learner) {Hint Learning\\Module};
    \node[database, above=0.8cm of learner] (offline_db) {Offline\\Labeled Data\\$(q, S, y^*)$};
    \node[draw=black, fill=white, rounded corners, below=0.6cm of learner, font=\scriptsize] (compiler_l) {Compiler};

    \node[agent, right=1.2cm of hintbank] (inference) {Hint-Guided\\Inference Module\\ \scriptsize \textit{(Logic Plan + Synthesis)}};
    \node[process, right=0.8cm of inference, fill=green!5, draw=green!40!black] (output) {Output\\SQL};
    \node[database, above=0.8cm of inference] (online_db) {Online\\Unlabeled Data\\$(q, S)$};
    \node[draw=black, fill=white, rounded corners, below=0.6cm of inference, font=\scriptsize] (compiler_r) {Compiler};

    \node[below=1.5cm of output] (human) {User Feedback};
    
    \node[agent, anchor=center] (sem_update) at (human -| hintbank) {Hint Learning Module\\\scriptsize \textit{(Asynchronous Update)}};

    
    \draw[->, thick] (offline_db) -- (learner);
    \draw[->, thick] (learner) -- (manager);
    \draw[->, thick] (manager) -- (hintbank);
    \draw[<->, thick] (learner) -- (compiler_l);

    \draw[->, thick] (online_db) -- (inference);
    \draw[->, thick] (hintbank) -- node[above, font=\scriptsize]{Retrieval} (inference);
    \draw[->, thick] (inference) -- (output);
    \draw[<->, thick] (inference) -- (compiler_r);

    \draw[->, thick] (output) -- (human);

    \draw[feedback] (human.west) -- (sem_update.east);

    \draw[feedback] (compiler_r.south) |- ([yshift=2mm]sem_update.east);
    
    \draw[feedback] (sem_update) -| (manager.south);

    \begin{scope}[on background layer]
        \coordinate (p1_tl) at ($(learner.west |- offline_db.north) + (-0.4, 0.6)$);
        \coordinate (p1_br) at ($(manager.west |- compiler_l.south) + (-0.3, -0.3)$);
        \draw[groupbox] (p1_tl) rectangle (p1_br);
        \node[label, anchor=north west] at (p1_tl) {Development Phase};

        \coordinate (p2_top_ref) at (online_db.north);
        \coordinate (p2_left_ref) at (inference.west);
        \coordinate (p2_right_ref) at (output.east); 
        \coordinate (p2_bottom_ref) at (human.south);

        \coordinate (p2_tl) at ($(p2_left_ref |- p2_top_ref) + (-0.3, 0.6)$);
        \coordinate (p2_br) at ($(p2_right_ref |- p2_bottom_ref) + (0.6, -0.6)$); 

        \draw[groupbox] (p2_tl) rectangle (p2_br);
        \node[label, anchor=north east] at ($(p2_br |- p2_tl)$) {Deployment Phase};
    \end{scope}

\end{tikzpicture}%
}
\caption{
The \sysname{} system architecture. 
The Hint Bank ($\mathcal{H}$) centrally manages both dialect-specific syntax hints and strategy-aware semantic hints across the Development and Deployment phases.
\textbf{Note:} This figure depicts the high-level \emph{data lifecycle} (how hints are generated, curated, and consumed); the detailed \emph{hint learning execution flow for a single example}---including multi-sampling, sequential syntax/semantic feedback, and multi-iteration refinement---is presented in Figure~\ref{fig:learning_process}.
}
\label{fig:arch}
\end{figure}
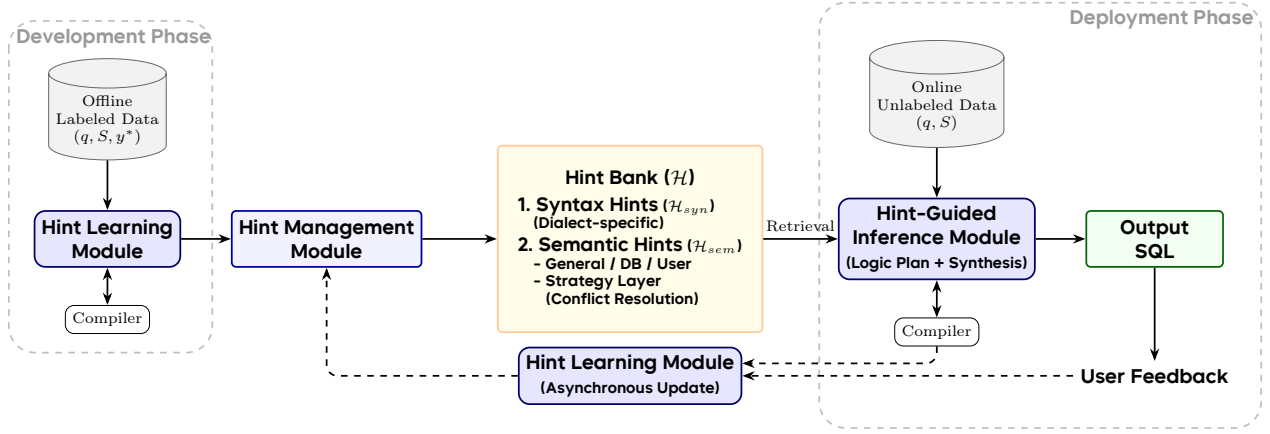

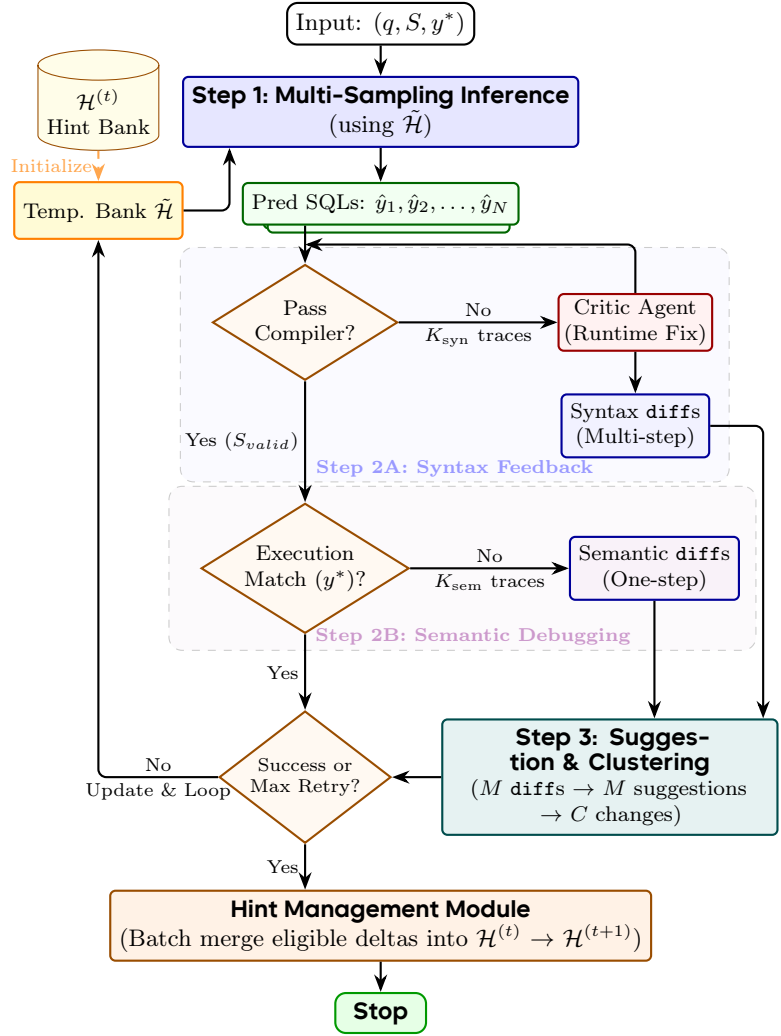
\begin{figure}[t]
\centering

\tikzset{
    apStartEndNode/.style={rectangle, draw=black, rounded corners, fill=white, thick, minimum width=1.6cm, align=center, font=\small},
    apProcessNode/.style={rectangle, draw=blue!60!black, fill=blue!5, thick, rounded corners=2pt, minimum height=0.75cm, align=center, font=\small},
    apDecisionNode/.style={diamond, draw=orange!60!black, fill=orange!5, thick, aspect=1.6, inner sep=1pt, font=\footnotesize, align=center},
    apContainer/.style={draw=gray!40, dashed, rounded corners=5pt, inner sep=6pt, fill=white},
    apArrow/.style={->, thick, rounded corners=3pt, >=Stealth}
}

\begin{tikzpicture}[
    node distance=0.95cm and 0.5cm,
    font=\sffamily\small
]

    \node[apStartEndNode] (input) {Input: $(q, S, y^*)$};

    \node[apProcessNode, below=0.4cm of input, minimum width=4.0cm, fill=blue!10] (inference)
    {\textbf{Step 1: Multi-Sampling Inference}\\(using $\tilde{\mathcal{H}}$)};
    \draw[apArrow] (input.south) -- (inference.north);

    \node[shape=cylinder, shape border rotate=90, draw=orange!60!black, fill=yellow!10,
          aspect=0.25, minimum height=0.8cm, minimum width=1.1cm, align=center,
          font=\footnotesize, left=0.3cm of inference] (hintbank) {$\mathcal{H}^{(t)}$\\Hint Bank};

    \node[apProcessNode,
          below=0.4cm of hintbank,
          fill=yellow!20, draw=orange,
          font=\footnotesize,
          align=center] (temphints) {Temp. Bank $\tilde{\mathcal{H}}$};

    \draw[apArrow, dashed, draw=orange!70]
        (hintbank.south) -- (temphints.north)
        node[midway, left, font=\scriptsize, inner sep=2pt, text=orange!70]
        {Initialize};  

    \draw[apArrow] (temphints.east) -| ([xshift=-2.0cm]inference.south);

    \node[apProcessNode,
          below=0.45cm of inference,
          fill=green!5, draw=green!40!black,
          font=\footnotesize,
          minimum width=3.2cm, minimum height=0.55cm,
          align=center] (pred_sql)
          {Pred SQLs: $\hat{y}_1, \hat{y}_2, \ldots, \hat{y}_N$};
    \draw[apArrow] (inference) -- (pred_sql);

    \node[apDecisionNode, below=0.5cm of pred_sql, xshift=-1.0cm] (check_syn) {Pass\\Compiler?};
    \draw[apArrow] ([xshift=-1.0cm]pred_sql.south) -- (check_syn);

    \node[apProcessNode,
          right=2.1cm of check_syn,
          align=center,
          font=\footnotesize,
          fill=red!5,
          draw=red!60!black,
          inner sep=2pt] (critic) {Critic Agent\\(Runtime Fix)};

    \draw[apArrow] (check_syn) --
      node[above, font=\scriptsize, inner sep=2pt] {No}
      node[below, font=\scriptsize, inner sep=2pt] {$K_{\text{syn}}$ traces}
      (critic);

    \node[apProcessNode,
          below=0.55cm of critic,
          fill=blue!5,
          font=\footnotesize,
          align=center,
          minimum width=1.9cm] (syn_diff) {Syntax \diff{}s\\(Multi-step)};
    \draw[apArrow] (critic) -- (syn_diff);

    \draw[apArrow] (critic.north) |- ([yshift=0.25cm]check_syn.north);

    \node[apDecisionNode, below=1.6cm of check_syn] (check_sem) {Execution\\Match ($y^*$)?};
    \draw[apArrow] (check_syn) -- node[left, font=\scriptsize, inner sep=2pt] {Yes ($S_{valid}$)} (check_sem);

    \node[apProcessNode,
          right=2.1cm of check_sem,
          fill=violet!5,
          font=\footnotesize,
          align=center,
          minimum width=1.9cm] (sem_diff) {Semantic \diff{}s\\(One-step)};

    \draw[apArrow] (check_sem) --
      node[above, font=\scriptsize, inner sep=2pt] {No}
      node[below, font=\scriptsize, inner sep=2pt] {$K_{\text{sem}}$ traces}
      (sem_diff);

    \node[apDecisionNode, below=1.0cm of check_sem, aspect=1.3, font=\scriptsize] (stop_check) {Success or\\Max Retry?};

    \node[apProcessNode,
      right=0.65cm of stop_check,   
      fill=teal!10, draw=teal!60!black,
      font=\footnotesize, align=center,
      text width=4.2cm] (suggestion)
      {\textbf{Step 3: Suggestion \& Clustering}\\
       ($M$ \diff{}s $\rightarrow$ $M$ suggestions\\
       $\rightarrow$ $C$ changes)};

    \coordinate (sem_target) at (suggestion.north-|sem_diff.south);
    \draw[apArrow] (sem_diff.south) -- (sem_target);
    
    \path (syn_diff.east) ++(0.7cm,0) coordinate (syn_out);
    \coordinate (syn_target) at (suggestion.north-|syn_out); 
    \draw[apArrow] (syn_diff.east) -- (syn_out) -- (syn_target);

    \draw[apArrow] (suggestion.west) -- (stop_check.east);
    \draw[apArrow] (check_sem.south) -- node[left, font=\scriptsize, inner sep=2pt] {Yes} (stop_check.north);

    \draw[apArrow] (stop_check.west) -| (temphints.south)
        node[near start, above, font=\scriptsize, inner sep=2pt] {No}
        node[near start, below, font=\scriptsize, inner sep=2pt] {Update \& Loop};

    \node[apProcessNode, below=0.6cm of stop_check, xshift=1.0cm, fill=orange!10, draw=orange!60!black,
          minimum width=5.5cm, align=center] (management)
          {\textbf{Hint Management Module}\\(Batch merge eligible deltas into $\mathcal{H}^{(t)}$ $\to$ $\mathcal{H}^{(t+1)}$)};
    \draw[apArrow] (stop_check.south) -- node[left, font=\scriptsize, inner sep=2pt] {Yes} ([xshift=-1.0cm]management.north);

    \node[apStartEndNode, below=0.4cm of management, fill=green!10, draw=green!60!black,
          minimum width=1.2cm, font=\small] (output) {\textbf{Stop}};
    \draw[apArrow] (management.south) -- (output.north);

    \begin{scope}[on background layer]
        \node[apProcessNode,
              fill=green!5, draw=green!40!black,
              font=\footnotesize,
              minimum width=3.2cm, minimum height=0.55cm,
              at={($(pred_sql.center)+(1.5pt,-1.5pt)$)}] {};
        \node[apProcessNode,
              fill=green!5, draw=green!40!black,
              font=\footnotesize,
              minimum width=3.2cm, minimum height=0.55cm,
              at={($(pred_sql.center)+(3pt,-3pt)$)}] {};

        \path (check_syn.west) ++(-0.2,0) coordinate (ref_left);
        \path (syn_diff.south) ++(0,-0.10) coordinate (syn_bottom_pad);
        \node[apContainer, fit=(check_syn) (critic) (syn_diff) (syn_bottom_pad) (ref_left), fill=blue!2] (syn_box) {};
        \node[anchor=south, font=\bfseries\scriptsize, text=blue!40, inner sep=2pt] at (syn_box.south)
            {Step 2A: Syntax Feedback};

        \path (check_sem.west) ++(-0.2,0) coordinate (sem_left_bound);
        \path (sem_diff.south) ++(0,-0.10) coordinate (sem_bottom_pad);
        \node[apContainer, fit=(check_sem) (sem_diff) (sem_left_bound) (sem_bottom_pad), fill=violet!2] (sem_box) {};
        \node[anchor=south, font=\bfseries\scriptsize, text=violet!40, inner sep=2pt, xshift=4pt] at (sem_box.south)
            {Step 2B: Semantic Debugging};
    \end{scope}

\end{tikzpicture}
\caption{The per-example hint learning loop. A temporary bank $\tilde{\mathcal{H}}$ is initialized from the frozen global bank $\mathcal{H}^{(t)}$; multi-sampled candidates undergo sequential \textbf{Syntax} and \textbf{Semantic} filtering, and the resulting atomic \diff{}s drive multi-iteration refinement of $\tilde{\mathcal{H}}$ until success or max retry. After all examples in the batch finish, the \textbf{Hint Management Module} consolidates the eligible temporary-bank deltas into $\mathcal{H}^{(t)}$ to produce $\mathcal{H}^{(t+1)}$.}
\label{fig:learning_process}
\end{figure}

Figure~\ref{fig:arch} illustrates the high-level architecture of \sysname{}.
Unlike traditional pipelines that treat prompts as transient strings, \sysname{} centers around a persistent, evolving knowledge base: the \textbf{Hint Bank ($\mathcal{H}$)}.
This centralized asset stores all learned \emph{Syntax Hints} ($\mathcal{H}_{syn}$) and \emph{Semantic Hints} ($\mathcal{H}_{sem}$).

The system operates in two distinct phases, designed to ``shift left'' the computational burden of error correction. The key distinction lies in the availability of ground truth:

\begin{itemize}
    \item \textbf{Phase I: Development Phase.} 
    The system bootstraps an initial $\mathcal{H}$ using a small offline dataset where \textbf{explicit Ground Truth SQL ($y^*$)} and \textbf{verified execution results} are available.
    By iteratively exposing the LLM to failures against $y^*$, the system extracts generalized $\mathcal{H}_{syn}$ and $\mathcal{H}_{sem}$ to \textbf{populate the Hint Bank}.
    This phase incurs high computational cost upfront to ensure efficiency later, and corresponds to the fully supervised learning regime detailed in Section~\ref{sec:learning_process}.
        
    \item \textbf{Phase II: Deployment Phase.} 
    Deployed in production, the system serves unseen queries and adapts via \textbf{asynchronous updates}.
    Syntax updates are automated via compiler feedback: failed online queries are batch-processed by the Syntax Learning Agent without blocking users.
    For semantic adaptation, the system relies on \emph{Human-in-the-Loop} interaction: each user-accepted SQL candidate is treated as a \textbf{pseudo-ground-truth ($\tilde{y}^*$)}, while a rejected candidate becomes a semantic error trace only when its execution result \emph{differs} from that of an accepted candidate; if all candidates are rejected, the query is logged for manual review rather than learned from.
    This two-stream update process ensures robustness to both dialect updates and shifting user intents.
    We detail this deployment lifecycle in Section~\ref{sec:deployment}.
\end{itemize}

\sysname{} comprises three core modules that manage the lifecycle of hints:

\noindent\textbf{1. Hint Learning Module (Section~\ref{sec:learning_process}).} \emph{The ``Writer.''} \\
\textbf{Input:} A query context $(q, S)$, the frozen global Hint Bank $\mathcal{H}^{(t)}$ for the current batch, and available feedback sources (the compiler, execution against $y^*$, or user feedback).\\
\textbf{Output:} Temporary-bank deltas (atomic \diff{}s clustered into suggestions and applied to a per-example $\tilde{\mathcal{H}}$), 
and, when the stop criteria retain them, the eligible temporary-bank deltas to be later consolidated by the Hint Management Module.\\
Given this input, the module invokes hint-guided inference to sample candidates and then converts observed failures into temporary-bank deltas. As detailed in Figure~\ref{fig:learning_process} and Section~\ref{sec:learning_process}, it implements a \textbf{three-stage learning pipeline}: 
(1) diversity via multi-sampling, 
(2) \emph{sequential} syntax--then--semantic feedback with error clustering and reflection, and 
(3) \emph{multi-iteration hint refinement} that re-injects updated hints to verify their effectiveness.
Concretely, it first interacts with a compiler to fix dialect-specific syntax errors (generating $\mathcal{H}_{syn}$), and subsequently compares execution results against the ground truth or user feedback to identify semantic mismatches (generating $\mathcal{H}_{sem}$).

\noindent\textbf{2. Hint Bank Management Module (Section~\ref{sec:management}).} \emph{The ``Curator.''} \\
\textbf{Input:} The eligible per-example temporary-bank deltas (from each $\tilde{\mathcal{H}}$) accumulated over a batch, together with the frozen global Hint Bank $\mathcal{H}^{(t)}$.\\
\textbf{Output:} The merged next-version global Hint Bank $\mathcal{H}^{(t+1)}$ (structured, deduplicated, and conflict-resolved).\\
After all examples in a batch finish, it consolidates their eligible deltas into $\mathcal{H}^{(t)}$ by merging newly learned hints with existing ones, deduplicating redundant triggers, and resolving conflicts.
A key innovation is the \textbf{Strategy Layer}, which models mutually exclusive behaviors (e.g., conflicting rounding conventions) as competing \emph{strategies} under a single trigger.
For each strategy, the module maintains lightweight signals---a learning-time recency signal and post-learning attribution statistics---that the retrieval and planning stages later use to rank strategies during inference and to balance ``correction'' behaviors against common ``vanilla'' behaviors learned from success cases. We defer the precise definition of these signals and the attribution procedure to Section~\ref{sec:management}.

\noindent\textbf{3. Hint-Guided Inference Module (Section~\ref{sec:inference}).} \emph{The ``Reader.''} \\
\textbf{Input:} A new user question $q$, schema $S$, and the Hint Bank $\mathcal{H}$.\\
\textbf{Output:} A ranked list of executable SQL candidates.\\
At inference time, it retrieves relevant $\mathcal{H}_{sem}$ via trigger matching and combines them with the complete syntax-hint registry $\mathcal{H}_{syn}^{(\text{dialect})}$ for the target dialect.
It then guides the base LLM through a \textbf{two-stage generation process} that mirrors Section~\ref{sec:inference}: \emph{Logic Planning} followed by \emph{SQL Synthesis}.
Crucially, the \textbf{Logic Planning} stage is designed to determine the optimal combination of strategies (especially when multiple retrieved hints offer competing paths) and to resolve conflicts using the statistics maintained by the Hint Bank Management Module, forming a coherent solution plan before the final SQL is synthesized.

\section{Methodology}
\label{sec:methodology}

\subsection{Data Model: Structure of Hints}
\label{sec:data_model}
To effectively serve as the system's long-term memory, the Hint Bank $\mathcal{H}$ employs an asymmetric design tailored to the distinct nature of the errors identified in Section 3.2.
We observe that \emph{syntax constraints} are objectively verifiable and static (determined by the compiler), whereas \emph{semantic logic} is often ambiguous and context-dependent (determined by users).
Consequently, we use a lightweight, rule-based structure for syntax hints, while employing a structured, conflict-aware schema for semantic hints to manage competing interpretations (see Section~\ref{sec:casestudy} for concrete structured hints paired with each walkthrough example).

\paragraph{Syntax Hints ($\mathcal{H}_{syn}$).}
These hints address dialect-specific compilation errors. Since syntax rules within a target dialect are deterministic and conflict-free (e.g., a keyword is either reserved or it is not), we do not require complex scoping or strategy selection. We structure each syntax hint simply as a tuple $(R, E)$:
\begin{itemize}
    \item \textbf{Rule ($R$):} A concise natural language description of the constraint (e.g., ``Quote every database element exactly as stored'').
    \item \textbf{Example ($E$):} A few-shot demonstration that grounds the rule, consisting of a specific schema context and the corresponding correct \emph{SQL snippet}.
\end{itemize}

\paragraph{Semantic Hints ($\mathcal{H}_{sem}$).}
These capture logic adjustments required by specific schemas or user preferences. Unlike syntax, semantic intents often carry ambiguity---for instance, a ``log transformation'' might imply different mathematical handling of zeros depending on the context. To ensure precise retrieval and explicit conflict resolution, we structure each semantic hint as a tuple $(T, S, \Sigma)$:

\begin{itemize}
    \item \textbf{Trigger ($T$):} A concise natural language pattern (e.g., ``log10 transformation'') used as the index key for retrieval. This decouples the hint from exact lexical matching, allowing the system to activate logic based on semantic similarity.
    
    \item \textbf{Scope ($S$):} A classification tag indicating applicability: \emph{General}, \emph{Database-specific}, or \emph{User-specific}. This design promotes knowledge transferability: general logic can be reused across projects, while database-specific conventions and user-specific preferences (e.g., tie-handling rules) remain isolated to their respective environments. We emphasize that \emph{General} here is purely a \emph{context} label: it states only that the hint is not tied to a specific database id or user id, in contrast to Database-specific and User-specific hints. It does \emph{not} claim that the hint is universally frequent, highly reliable, or broadly useful across all workloads. Reliability and usefulness are tracked separately, after learning, by the Strategy Attribution pass through $\sigma.\texttt{eval\_stats}$ (Section~\ref{sec:management}); two General hints with very different empirical support can therefore coexist and be ranked accordingly at inference time.

    \item \textbf{Strategies ($\Sigma$):} A set of competing strategies $\{\sigma_1, \dots, \sigma_k\}$ associated with the trigger. This layer is critical for handling conflicts. For example, the trigger ``log10 transformation'' may map to two valid but mutually exclusive strategies: \emph{Strategy A} (add 1 to avoid infinity) and \emph{Strategy B} (nullify zeros).
    Crucially, each strategy $\sigma_i$ contains:
    (1) a \emph{rationale},
    (2) a pair of \emph{contrastive few-shot examples} (Positive vs.\ Negative) to distinguish the logic, and
    (3) two pieces of \emph{statistical metadata}: the \emph{learning-time} field \texttt{recency}, written by the Hint Learning Module to record when the strategy was last distilled, and the \emph{evaluation-time} field $\texttt{eval\_stats}$, written exclusively by the post-learning Strategy Attribution pass (Section~\ref{sec:management}) and used to quantify the strategy's empirical effect on actual generations. The two fields are updated by disjoint mechanisms and never overwrite each other.
\end{itemize}

\subsection{Hint Learning Module}
\label{sec:learning_process}
Instead of treating prompt optimization as a one-off search, \sysname{} adopts an iterative, debugging-based approach. The core philosophy follows a standard engineering cycle: \textbf{Execution $\rightarrow$ Feedback $\rightarrow$ Reflection $\rightarrow$ Re-evaluation}. 
For each example, \sysname{} does not directly modify the global Hint Bank. Instead, it initializes a \textbf{temporary Hint Bank} $\tilde{\mathcal{H}}$ as a working copy of the current frozen global bank $\mathcal{H}^{(t)}$, and all multi-iteration revisions are applied exclusively to $\tilde{\mathcal{H}}$. This design keeps the global bank stable during iterative refinement. Importantly, the per-example output is \emph{not} a new global bank: each example produces only a set of \textbf{temporary-bank deltas}---the additions, revisions, and suspensions recorded in its own $\tilde{\mathcal{H}}$ relative to $\mathcal{H}^{(t)}$. After \emph{all examples in the current batch} finish their per-example loops, the Hint Management Module (Section~\ref{sec:management}) merges the eligible temporary-bank deltas into the frozen $\mathcal{H}^{(t)}$ to produce the next version $\mathcal{H}^{(t+1)}$, following the batch-sequential protocol detailed in Section~\ref{sec:deployment}.

The remainder of this subsection zooms in on the \textbf{per-example} learning loop: the three integrated stages 
below all operate on a single example and its temporary Hint Bank $\tilde{\mathcal{H}}$, and they repeat until a stop 
criterion is met.

\subsubsection{Step 1: Multi-Sampling Inference with the Temporary Hint Bank}
Single-pass generation is inherently noisy due to LLM stochasticity. To ensure that revisions to $\tilde{\mathcal{H}}$ cover the 
full spectrum of potential errors and do not introduce regressions, we employ \textbf{Multi-Sampling} at the start of each 
iteration.
For a given input $(q, S)$, we perform $N$ inference runs via the \textbf{Hint-Guided Inference Module} 
(Section~\ref{sec:inference}), each conditioned on the \textbf{temporary Hint Bank} 
$\tilde{\mathcal{H}}$ (initialized from $\mathcal{H}^{(t)}$, with $\mathcal{H}^{(0)}=\emptyset$ for the first batch). 
The temporary-bank deltas accumulated in $\tilde{\mathcal{H}}$ are not committed to the global bank one example at a time; 
they are pooled across the current batch and consolidated into $\mathcal{H}^{(t+1)}$ collectively 
(see~\S\ref{sec:deployment}).

\subsubsection{Step 2: Sequential Feedback \& Atomic \diff{} Generation}
We process the $N$ generated candidates (i.e., samples) through a two-stage sequential filter, using the ground-truth execution 
result as the reference. In the first stage, candidates that fail to compile form the set of \emph{syntax error traces}, 
denoted by $K_{\text{syn}}$. We then apply runtime syntax fixes (Phase~A) to obtain syntactically valid SQL queries, 
including both originally valid and repaired candidates. All syntactically valid candidates are subsequently executed, 
and those whose execution results still fail to match the ground truth form the set of \emph{semantic error traces}, 
denoted by $K_{\text{sem}}$.

\paragraph{Phase A: Syntax Correction \& \diff{} Generation (Compiler Feedback).}
The first barrier is the SQL compiler. For each candidate in $K_{\text{syn}}$, we enter a rapid \textbf{Critic Loop} that asks 
an agent to repair syntax without altering semantics. Because this loop involves iterative interaction with compiler feedback, 
syntax correction is inherently a \textbf{multi-step} process. From the resulting sequence of compiler errors and repairs, 
we extract atomic syntax \diff{}s that capture specific violations and their corresponding fixes.
We extract atomic syntax \diff{}s \emph{only} from candidates whose Critic Loop ultimately produces a syntactically valid query: 
a successful repair trace gives an unambiguous (error~$\rightarrow$~fix) pair from which a hint can be distilled. 
Candidates that exhaust the Critic Loop's iteration budget without compiling are discarded for the purpose of \emph{syntax} hint 
learning, since their final state offers no verified fix to generalize from.

\paragraph{Phase B: Semantic Debugging \& \diff{} Generation (Execution Feedback).}
Semantic debugging proceeds in two steps. First, the execution-result mismatch (against the ground-truth result) identifies \emph{which} candidates are semantically wrong, yielding $K_{\text{sem}}$. Then, since offline learning has access to the ground-truth SQL $y^*$, a semantic \diff{} agent performs a \textbf{one-step} comparison between the \emph{logical steps} of $y^*$ and the predicted SQL (optionally aided by their result summaries) to localize the specific logic gaps responsible for the mismatch (e.g., an incorrect join condition or aggregation scope).

To make the feedback produced by these two phases comparable and reusable, \sysname{} represents both syntax and semantic deviations as \textbf{atomic \diffs{}}. An atomic \diff{} isolates a single, specific deviation between the preferred solution and an error trace, avoiding the complexity of comparing full SQL queries or verbose execution logs directly. The unified schema for atomic \diffs{} is detailed in Appendix~\ref{app:hint_diffs}.

\subsubsection{Step 3: Hint Suggestion \& Clustering (Updating the Temporary Hint Bank)}
\label{sec:multi_iteration}
After processing the $K_{\text{syn}}$ syntax error traces and $K_{\text{sem}}$ semantic error traces, we obtain a set of $M$ atomic \diff{}s (from both compiler and execution feedback). 
We then prompt another LLM to produce a \textbf{hint suggestion} for each \diff{}. Specifically, the agent must decide to:
(1) \textbf{Add a new hint}, if no hint in the current $\tilde{\mathcal{H}}$ would have prevented the error described by the \diff{}; or
(2) \textbf{Revise an existing hint}, if a hint already in $\tilde{\mathcal{H}}$ \emph{should} have prevented the error but failed (e.g., due to ambiguity), and thus needs refinement; or
(3) \textbf{Suspend a previously proposed change}, i.e., roll back an addition or revision \emph{introduced earlier in the current temporary-bank refinement loop} if the latest \diff{} indicates that the change is now misleading or no longer applies---this lets the loop undo its own recent updates rather than only growing $\tilde{\mathcal{H}}$ monotonically. 

This yields $M$ hint suggestions. Because different traces may trigger overlapping or duplicate suggestions, we prompt an LLM to \textbf{cluster} these $M$ suggestions into $C$ final changes. These $C$ changes are then applied to update $\tilde{\mathcal{H}}$.
Each suggested semantic hint is also assigned a scope tag $S$ at generation time based on the current example's DB/User context, following the scope rules in Section~\ref{sec:data_model}.
The loop then restarts at Step~1, leveraging the refined $\tilde{\mathcal{H}}$. The core intuition is that complex traces are difficult to analyze holistically; the pipeline of \textbf{atomic \diffs{} $\rightarrow$ suggestions $\rightarrow$ clustering} distills actionable, empirically effective hints.

\paragraph{Stop Criteria and Conditional Merge.}
The hint refinement loop terminates under one of two conditions, each triggering a different merge policy:
\begin{enumerate}
    \item \textbf{All samples correct (Success).} All $N$ candidates generated with the current $\tilde{\mathcal{H}}$ pass both the compiler and execution checks against $y^*$. This confirms that the refined hints are strictly beneficial, so the temporary-bank deltas are \textbf{unconditionally} marked eligible for batch merge by the Hint Management Module (Section~\ref{sec:management}).

    \item \textbf{Maximum iterations reached.} The loop exhausts its revision budget $T_{\max}$ without achieving full correctness. To guard against regressions introduced by over-revision, we apply a \textbf{conservative merge policy}: the temporary-bank deltas are retained for batch merge \emph{only if} the number of correct samples produced under $\tilde{\mathcal{H}}$ strictly exceeds that of the original frozen bank, i.e., $|\mathrm{correct}(\tilde{\mathcal{H}})| > |\mathrm{correct}(\mathcal{H}^{(t)})|$. Both counts are measured under the \emph{same} sampling budget ($N$ samples) and decoding setting, so the comparison reflects the effect of the revised hints rather than sampling noise. If not, the deltas are discarded and do not enter the batch-level merge, preventing the global Hint Bank from being polluted by ineffective revisions.
\end{enumerate}

\subsection{Hint Management Module}
\label{sec:management}
A key challenge in automated hint learning is handling conflicts and maintaining distribution balance. If the system learns solely from errors, the Hint Bank risks becoming a collection of edge cases, losing track of general correct behaviors. We propose a management pipeline to ensure both consistency and coverage:

\begin{enumerate}
    \item \textbf{LLM-Augmented Merging \& Deduplication (Scope-Aware):} 
    After all examples in a batch finish, the Hint Management Module consolidates their eligible temporary-bank deltas into the global Hint Bank $\mathcal{H}$. 
    For each new or revised hint, we first determine its scope tag $S$ (Section~\ref{sec:data_model}) and only compare it against existing global hints with the same scope. 
    Within that scope, we check whether a similar \emph{Trigger} already exists using a \emph{Hint-Merge Agent}. 
    If the temporary hint matches an existing strategy, we merge them (e.g., consolidating examples and updating statistics). 
    If it proposes conflicting logic (e.g., Round-Up vs.\ Round-Down), the agent appends it as a new, mutually exclusive \emph{Strategy} under the same trigger, enabling the system to model localized variations without polluting other scopes.
    
    \item \textbf{Coverage Assurance (Preventing Error-Bias):} To prevent the Hint Bank from biasing towards failures (``overfitting to mistakes''), we additionally process examples that are solved \emph{on the first iteration}---i.e., where all $N$ candidates produced from the very first Multi-Sampling pass already pass both compiler and execution checks, so no error trace is generated. For such examples we still inspect the retrieved hints used during that first pass: an LLM judge compares the successful predicted SQL against the retrieved hints (via the same logical-\diff{} analysis used for semantic debugging) to infer the strategy actually used. If the model succeeded via parametric knowledge but the retrieved hints did not explicitly cover that strategy, we distill this \textbf{Vanilla Strategy} and add it to the Hint Bank. This ensures that ``normal'' or ``correct'' behaviors are competing fairly with ``correction'' strategies during ranking, instead of being silently absent from the bank simply because they never produced an error.
    
    \item \textbf{Strategy Attribution and Statistic Tracking:} Each strategy $\sigma$ carries two complementary signals. The first is a \emph{learning-time} signal, \texttt{recency}: the timestamp of the most recent successful application during learning, used by the retriever to prioritize evolving user preferences. We deliberately do \emph{not} rely on raw learning-time success counts, because they conflate ``how often a strategy was \emph{learned}'' with ``how often it was \emph{useful}'', which inflates recall but hurts retrieval precision once the bank grows.

    The second signal is an \emph{evaluation-time} attribution measured on a \emph{frozen} Hint Bank. Periodically---e.g., after every $K$ learned examples or at the end of a learning round---we pause learning, freeze the current global Hint Bank $\mathcal{H}$, and run a dedicated multi-sample evaluation pass on previously processed examples using $\mathcal{H}$ as the only knowledge source. This evaluation produces, for each example, the predicted SQL, the ground-truth SQL, the execution score, and the set of retrieved hint IDs; the resulting artifacts are then consumed by the \textbf{Strategy Attribution} step, which writes its results into a dedicated field $\sigma.\texttt{eval\_stats}$ that is never modified by the learning loop. Because retrieval is recorded at the hint level but a retrieved hint exposes all of its strategies, the exposed strategy set for a sample is the union of strategies under its retrieved hints. For every (example, exposed strategy) pair, an LLM judge inspects the predicted SQL against the ground-truth SQL and assigns one of three SQL-evidence-driven verdicts:
    \begin{itemize}
        \item \emph{positive}: the predicted SQL visibly applied $\sigma$'s preferred action and that choice was relevant to a correct execution result;
        \item \emph{negative}: the predicted SQL followed $\sigma$'s preferred action and that exact choice is where it diverged from the ground-truth SQL;
        \item \emph{inert}: $\sigma$ was retrieved but not reflected in the SQL, or its application was irrelevant to correctness.
    \end{itemize}
    Verdicts are aggregated to the example level across the $N$ samples: an example is added to a strategy's \emph{positive} list if any correct sample applied $\sigma$, and to its \emph{negative} list if any incorrect sample was misled by $\sigma$; inert exposures in incorrect samples are \emph{not} counted as negative.
    Because attribution is aggregated across multiple samples, the positive and negative lists are not necessarily mutually exclusive: the same example may show that a strategy can help one sample while misleading another. We therefore do \emph{not} force a single label per (example, strategy) pair; instead, the inference-time formatter (Section~\ref{sec:inference}) reports success rate and harm rate as separate quantities over the retrieved support, so the planner can see partial-help cases for what they are.
    Concretely, $\texttt{eval\_stats}$ keeps three deduplicated lists per strategy---the examples on which it was retrieved, those it helped, and those it hurt---over the previously evaluated examples; storing the actual example references (rather than only counts) keeps the data deduplicated, auditable, and recomputable across attribution runs.
\end{enumerate}

\subsection{Hint-Guided Inference Module}
\label{sec:inference}
During inference, \sysname{} utilizes the Hint Bank to guide the LLM. The process proceeds in two sequential stages:

\begin{enumerate}
    \item \textbf{Step 1: Hint Retrieval and Logic Planning.}
    Before generating the SQL query, we retrieve relevant semantic hints from the Hint Bank in two stages. We first apply a \emph{scope-aware filter} (Section~\ref{sec:data_model}) over the global bank: \emph{General} hints are always kept, while \emph{Database-specific} and \emph{User-specific} hints are kept only when their attached database id (resp.\ user id) matches the current query's context. Scope filtering is performed \emph{before} retrieval, so the candidate pool exposed to retrieval is already context-isolated.
    The remaining candidates' \emph{Triggers} (and aliases) are then serialized into a compact textual list and passed to a \emph{retrieval LLM} (run with low temperature), which is instructed to use semantic judgment rather than keyword matching to return the subset of hint ids it considers relevant to $q$.
    The associated strategies are scored using the flexible signals defined in Section~\ref{sec:management}: the learning-time \texttt{recency} (used to prioritize recent user contexts) and the evaluation-time attribution stored in $\sigma.\texttt{eval\_stats}$.
    Crucially, before any of this is exposed to the planning LLM we apply a lightweight \textbf{formatting layer} that turns the raw \texttt{eval\_stats} lists into compact, anonymized statistics: it computes a per-strategy \emph{success rate} (positive over retrieved), the corresponding \emph{harm rate} (negative over retrieved), the inert share, and the support size, and drops the underlying example references. This serves two purposes simultaneously: it prevents leakage of any specific evaluation example into the prompt, and it presents the planning LLM with a single trustworthiness summary per strategy rather than raw bookkeeping.
    The LLM is then asked to use these summaries as a credibility signal: a strategy with a high success rate over many distinct examples is treated as well-supported and preferred; one that is frequently retrieved but rarely actually applied (high inert share) is surfaced as optional and may be safely ignored; one with a non-trivial harm rate or only a handful of supporting examples is flagged as low-confidence and can be down-weighted or skipped. The LLM then produces a \emph{Logic Plan} that integrates the strategies it chooses to trust, resolving any remaining conflicts before code generation.
    
    \item \textbf{Step 2: SQL Synthesis.} Following logic planning, the system injects the \textbf{complete set} of dialect-specific \emph{Syntax Hints} alongside the generated Logic Plan into the prompt for the final SQL synthesis. 
    We consciously inject \emph{all} available syntax hints for the target dialect rather than retrieving a subset. Because modern base LLMs already possess a strong foundational understanding of SQL, the number of persistent, dialect-specific compilation errors tends to be small. Consequently, the syntax registry for any single dialect remains highly compact. Injecting the complete set provides exhaustive dialect guidance and substantially improves adherence to compiler constraints without risking prompt context overflow, allowing the LLM to reliably synthesize the final executable SQL.
\end{enumerate}

\subsection{Deployment Lifecycle}
\label{sec:deployment}
\sysname{} is designed for continuous adaptation, bridging the gap between static training and dynamic production through a two-phase lifecycle.

\paragraph{Phase I: Development (Cold Start).}
We bootstrap the initial Hint Bank $\mathcal{H}$ using a small, manually curated offline dataset.
Crucially, each example in this phase includes not just the ground truth SQL ($y^*$), but also the verified \textbf{Execution Result}, allowing both the Syntax and Semantic Learning Agents to operate in full capacity: validating syntax against the compiler and verifying logic by comparing execution outcomes against~$y^*$.

\textbf{Batch-Sequential Update Protocol.}
The offline dataset is processed in sequential batches $B_1, B_2, \ldots, B_T$.
We denote the Hint Bank available at the start of batch $B_t$ as $\mathcal{H}^{(t)}$, with $\mathcal{H}^{(0)} = \emptyset$ (empty bank).
For each example in $B_t$, the Hint Learning Module executes the full per-example pipeline (Figure~\ref{fig:learning_process}): a temporary Hint Bank $\tilde{\mathcal{H}}$ is initialized as a copy of the frozen $\mathcal{H}^{(t)}$, and all multi-iteration hint revisions operate exclusively on $\tilde{\mathcal{H}}$.
Because all examples within $B_t$ share the same frozen $\mathcal{H}^{(t)}$, their per-example $\tilde{\mathcal{H}}$ copies are independent and can be processed in parallel.
Once all examples in $B_t$ are processed, the Hint Management Module merges all eligible temporary-bank deltas into $\mathcal{H}^{(t)}$ to produce the updated bank $\mathcal{H}^{(t+1)}$. This batch merge is the only point at which the global bank is updated.
This sequential batch structure ensures every batch builds cumulatively on all knowledge distilled by prior batches.
After $T$ batches, the resulting $\mathcal{H}^{(T)}$ serves as the warm-started Hint Bank handed off to the Deployment Phase.

\paragraph{Phase II: Deployment.}
The fundamental challenge distinguishing deployment from the development phase is the \textbf{absence of ground truth SQL}. 
In development, we have access to a verified $y^*$ for every example, which enables both the Syntax and Semantic Learning Agents to operate at full capacity.
In production, no such ground truth is predefined, so the two types of hint learning require different adaptation strategies.

\begin{itemize}
    \item \textbf{Continuous Syntax Updates (Automated):} Syntax learning is unaffected by the absence of ground truth SQL, because the SQL compiler itself is the sole source of truth for syntax validity.
    Failed queries from live traffic are batch-processed asynchronously by the \emph{Syntax Learning Agent} exactly as in the development phase, allowing the system to adapt to dialect version updates or newly encountered errors without any human intervention.

    \item \textbf{Human-in-the-Loop Semantic Updates:} Semantic learning requires knowing which SQL output is \emph{correct}, which is precisely what ground truth provides in development.
    In deployment we approximate this via a \textbf{user preference interaction}.
    For each incoming question $q$, the system runs $N$ inference calls with the current Hint Bank, generating $N$ candidate SQL queries $\{\hat{y}_1, \ldots, \hat{y}_N\}$ ranked by strategy statistics (Section~\ref{sec:management}).
    These candidates are presented to the user, who \textbf{accepts} any that produce a correct result and \textbf{rejects} the rest.

    This labeling induces a \emph{preference dataset} analogous to the signal used in Direct Preference Optimization (DPO)~\cite{rafailov2023direct}:
    \begin{enumerate}
        \item \textbf{Positive signal (accepted):} Each accepted candidate is treated as a pseudo-ground-truth $\tilde{y}^*$.
        \item \textbf{Negative signal (rejected):} Each rejected candidate $\hat{y}_{\text{rej}}$ whose execution result \emph{differs} from an accepted one is treated as a semantic error.
        \item \textbf{Hint learning:} The pair $(\tilde{y}^*, \hat{y}_{\text{rej}})$ is fed directly into the \emph{Semantic Learning Agent}, which generates atomic \diff{}s, produces hint suggestions, and updates $\mathcal{H}_{sem}$ via the Hint Management Module---exactly as in the development phase.
    \end{enumerate}
    If the user \textbf{rejects all} candidates, the system logs the query for manual review, as this signals a knowledge gap that cannot be safely resolved by automated reflection alone.
\end{itemize}

\subsection{Case Study: End-to-End Hint Learning}
\label{sec:casestudy}

We trace three concrete examples through the learning pipeline; the corresponding structured hints appear immediately after each example below.
Recall the three steps in each iteration: \textbf{Multi-Sampling Inference} (generating $N$ candidates from the current $\tilde{\mathcal{H}}$), \textbf{Sequential Feedback} (compiler and execution checks that produce atomic \diff{}s), and \textbf{Hint Suggestion \& Clustering} (converting \diff{}s into hint updates for $\tilde{\mathcal{H}}$).
Examples~1--2 walk through both feedback channels in full; Example~3 illustrates a \emph{Database-specific} semantic hint. (No \emph{User-specific} examples appear here, since this benchmark exposes no per-user identity; see Section~\ref{sec:experimental_setup}.)

\paragraph{Example 1 (Syntax): Snowflake Identifier Quoting.}
Consider the question \emph{``Count all orders placed''} against an orders table.
In the \textbf{Multi-Sampling Inference} step with an empty $\tilde{\mathcal{H}}$, all sampled candidates use bare, unquoted identifiers (e.g., \texttt{FROM SALES.ORDERS}).
During \textbf{Sequential Feedback}, the compiler rejects every candidate, reporting that the table object does not exist---Snowflake requires every identifier to be double-quoted in the exact stored casing.
This compiler trace is converted into a \texttt{SYNTAX}-category atomic \diff{} whose gold strategy is ``quote each path element separately in double-quotes''.
In the \textbf{Hint Suggestion \& Clustering} step, this \diff{} is translated into a syntax hint---a rule stating that every identifier must be quoted exactly as stored, accompanied by a corrected SQL snippet---and added to $\tilde{\mathcal{H}}_{syn}$.
In the next iteration's Multi-Sampling Inference step, the hint is injected into the prompt and all candidates now compile successfully.
The success stop criterion is met, and the resulting temporary-bank delta is marked eligible for the batch-level merge.

\begin{tcolorbox}[colback=blue!5!white, colframe=blue!75!black, title=\textbf{Structured hint (Example~1)}]
\textbf{Rule:} Quote every database, schema, table, column, CTE, and alias exactly as it is stored; quote each element of a fully-qualified path separately. System-generated columns from table-functions are already UPPER-CASE---leave them unquoted or quote them in UPPER-CASE only.

\textbf{Example:}
Schema: \texttt{sales.orders(orderId, order\_date)}
Question: \texttt{Count all orders.}

\begin{lstlisting}[language=SQL,basicstyle=\ttfamily\small]
SELECT COUNT(*) AS "total"
FROM "SALES"."ORDERS";
\end{lstlisting}
\end{tcolorbox}

\paragraph{Example 2 (Semantic, General): log$_{10}$ Transformation.}
For the question \emph{``Compute the log\textsubscript{10}-transformed view count per event type''}, the \textbf{Multi-Sampling Inference} step produces candidates that use \texttt{NULLIF(col, 0)} to suppress zeros before taking the log.
The queries compile, but during \textbf{Sequential Feedback} (execution phase) the result set mismatches the ground truth: rows with a zero view count return \texttt{NULL} instead of the expected \texttt{0} (i.e., $\log_{10}(0 + 1)$).
The LLM analyzes the ground-truth SQL versus the predicted output and extracts a \texttt{FORMULA}-category \diff{}: the gold strategy adds 1 before the log to handle zero-valued rows, while the wrong strategy silently nullifies them.
\textbf{Hint Suggestion \& Clustering} converts this into a general-scope semantic hint with trigger ``log10 transformation of counts'', containing a strategy that prescribes adding one before applying log10, i.e., $\log_{10}(\text{column} + 1)$, and flags nullifying zeros (e.g., \texttt{NULLIF}) as incorrect.
Because the scope is \emph{General}, this hint is eligible for retrieval across databases and users; whether it should be trusted for a specific query is still determined by attribution statistics during inference.
In the next iteration the retrieved hint guides the Logic Plan, and the corrected formula is used, matching the ground truth.

\begin{tcolorbox}[colback=blue!5!white, colframe=blue!75!black, title=\textbf{Structured hint (Example~2)}]
\textbf{Trigger:} \texttt{log10 transformation of counts}

\textbf{Strategy 1}

\textbf{Rationale:}
When the question asks for a log10 transformation of count-type data that can contain zeros, keep zero-valued rows and avoid $-\infty$ by adding 1 before the log.

\textbf{Preferred Action (logical):}
Apply $\log_{10}$ after adding one to the column: $\log_{10}(\text{column} + 1)$.

\textbf{Implementation example (Snowflake):}
\begin{lstlisting}[language=SQL,basicstyle=\ttfamily\small]
LOG(10, "{COLUMN}" + 1)
\end{lstlisting}

\textbf{Wrong Action (logical):}
Replace zeros with NULL before taking the log, which silently drops the zero-valued rows.

\textbf{Implementation example of the wrong action (Snowflake):}
\begin{lstlisting}[language=SQL,basicstyle=\ttfamily\small]
LOG(10, NULLIF("{COLUMN}", 0))
\end{lstlisting}
\end{tcolorbox}

\paragraph{Example 3 (Semantic, DB-specific): GA4 Visitor ID.}
For the question \emph{``List unique visitors from last month''} on the GA4 database, sampled candidates filter on the \texttt{USER\_ID} column, which is almost always \texttt{NULL} in the GA4 sample schema, producing an empty result.
The execution feedback identifies the correct column as \texttt{USER\_PSEUDO\_ID} (which stores the actual visitor identifier).
The learned hint is scoped to \emph{DB-specific (GA4)}, so it is only retrieved for queries against the GA4 database and does not affect other schemas.

\begin{tcolorbox}[colback=blue!5!white, colframe=blue!75!black, title=\textbf{Structured hint (Example~3)}]
\textbf{Database:} \texttt{GA4} \quad \textbf{Trigger:} reference to column \texttt{USER\_ID} in GA4 events tables

\textbf{Strategy 1}

\textbf{Rationale:}
GA4 sample ecommerce tables store visitor IDs in \texttt{USER\_PSEUDO\_ID}, while \texttt{USER\_ID} is almost always \texttt{NULL}.

\textbf{Preferred Action:}
Use the pseudo-ID column:
\begin{lstlisting}[language=SQL,basicstyle=\ttfamily\small]
... WHERE USER_PSEUDO_ID LIKE '{INT}.%' ...
\end{lstlisting}

\textbf{Wrong Action:}
Avoid filtering directly on \texttt{USER\_ID}:
\begin{lstlisting}[language=SQL,basicstyle=\ttfamily\small]
... WHERE USER_ID = '{INT}' ...
\end{lstlisting}
\end{tcolorbox}

\section{Evaluation}
\label{sec:evaluation}

\subsection{Experimental Setup}
\label{sec:experimental_setup}
We evaluate \sysname{} on the \textbf{Spider~2.0--Snow-0212} benchmark~\cite{spider2025snow}, which is designed to mimic realistic enterprise challenges, featuring the strict \textbf{Snowflake SQL dialect}, complex multi-table schemas, and logic-heavy questions that frequently confuse standard models.

\paragraph{Dataset Split.}
This paper focuses on the \emph{Development Phase} (Phase~I) of \sysname{}: hint learning and evaluation on examples that ship with usable supervision. The Spider~2.0--Snow-0212 release contains \textbf{547 examples} in total, of which only a subset ships with both ground-truth SQL and ground-truth execution results. After filtering, we obtain \textbf{113 examples} with usable supervision\footnote{Of the 547 examples, 120 are released with both a ground-truth SQL and a ground-truth execution result. We discard 7 of these whose official ground-truth SQL produces a result set that does not match the official ground-truth result when executed under the Snowflake engine, and treat them as having no usable ground truth. This leaves 113 examples with usable supervision (used for both hint learning and evaluation); the remaining $547-113=434$ examples do not have usable gold SQL for hint learning, but they do provide official execution-result targets, which we use only for evaluation. These 434 examples form the held-out split used in Section~\ref{sec:eval_heldout}.}\!, on which the Hint Bank is bootstrapped sequentially. Following the development-phase protocol described in Section~\ref{sec:deployment}, we then re-evaluate on the same 113 examples using the \emph{final} learned Hint Bank and compare against a vanilla baseline that uses no hints. We additionally evaluate, in Section~\ref{sec:eval_heldout}, on the remaining Spider~2.0--Snow-0212 examples that the development phase did \emph{not} learn from, in order to test how the development-phase Hint Bank generalizes outside its learning set. The full \emph{Deployment Phase} workflow---in which the same architecture continues to learn hints from these examples through accumulated execution and user feedback rather than ground-truth SQL---is left for future work (Section~\ref{sec:discussion}).

\paragraph{Models.}
We deliberately span three backbones of different capability tiers to test both end-to-end performance and \emph{cross-model transferability} of the learned Hint Bank:
\begin{itemize}
    \item \textbf{Doubao-2.0-lite} (weak): a lightweight open-style backbone used to test whether a Hint Bank distilled by a stronger model can lift a weaker generator.
    \item \textbf{GPT-5} (medium): a mid-tier OpenAI model used as a medium-capability transfer target.
    \item \textbf{GPT-5.5} (strong, \emph{primary}): our \emph{core} model. All learning agents (Syntax/Semantic Learning, Hint-Merge, Strategy Attribution) and the primary Hint Bank reported in Tables~\ref{tab:main_results} and~\ref{tab:ablation_attr} use GPT-5.5.
\end{itemize}
Unless otherwise stated, the Hint Bank evaluated in every table is the one learned by GPT-5.5 on the 113 development examples; Doubao and GPT-5 inherit \emph{exactly} this bank without any model-specific adaptation. The final bank contains \textbf{11 syntax hints} and \textbf{37 semantic hints} (\textbf{19} \emph{General} + \textbf{18} \emph{Database-specific}; no \emph{User-specific} hints arise in this single-tenant benchmark).

\paragraph{Baselines.}
For every backbone, our primary baseline is the \textbf{Vanilla} configuration of the same model: same backbone, decoding parameters, and syntax critic budget, but without hint retrieval or hint-conditioned planning. This isolates the contribution of the Hint Bank from raw model capability.

In addition, on the held-out split (Section~\ref{sec:eval_heldout}) we compare against a \textbf{reproduced SQLGenie-style Example-Bank RAG} baseline based on \emph{SQLGenie}~\cite{ghosh2025sqlgenie}, the closest published RAG-style system from our related work (Section~\ref{sec:related_work}). Since SQLGenie's official inference code is not available, we implement a best-effort reproduction of \emph{only} its Example Bank RAG route, omitting non-RAG components such as table onboarding, self-refinement, and feedback-driven bank augmentation. The bank stores the same 113 supervised development question--SQL pairs used by \sysname{} for hint learning; retrieval uses masked-question similarity and, for each query, returns the top-3 most similar (question, SQL) pairs, which are injected verbatim into the generator prompt. Evaluation is restricted to held-out examples whose gold SQL is not in the bank to avoid target leakage. This baseline (referred to as \textbf{SQLGenie-style RAG}) is run with \textbf{GPT-5.5} as the generator, with the same decoding parameters as the Vanilla and \sysname{} configurations.

\paragraph{Metrics.}
We report four metrics that jointly capture end-to-end correctness and the efficiency of iterative refinement:
\begin{itemize}
    \item \textbf{Pass Rate:} the fraction of \emph{sampled candidates} (across all examples) whose execution result matches the ground truth. We deliberately report this as a per-candidate rate because it is computed over all $k$ samples per example, not just a single top-ranked candidate.
    \item \textbf{pass@4 (pass@k):} an example counts as solved if at least one of the $k{=}4$ parallel candidates matches the ground truth.
    \item \textbf{Syntax Pass Rate:} the fraction of \emph{sampled candidates} that become syntactically valid Snowflake queries after the syntax critic loop terminates (i.e., either parsed successfully on initial generation or repaired into valid SQL within the per-sample critic budget). We report this at the candidate level for consistency with Avg.\ Critics, which is also aggregated over candidates.
    \item \textbf{Avg.\ Critics:} the average number of compiler-feedback critic rounds taken across all sampled candidates, where each candidate's count is measured until it either becomes syntactically valid or exhausts the per-sample critic budget. This is an aggregate over candidates, not an example-level ``first valid candidate'' metric. Lower is better.
\end{itemize}

\paragraph{Implementation Details.}
All experiments are executed under the Snowflake dialect of Spider~2.0--Snow-0212. For \sysname{}'s \emph{Syntax Loop}, we use the Snowflake SQL compiler for real-time validation; for the \emph{Semantic Loop}, correctness is measured by comparing execution result sets against the ground truth. We set the maximum number of hint-learning iterations per example to \textbf{3}, draw \textbf{$k{=}4$} candidate samples in parallel at inference time, and use a sampling temperature of \textbf{0.3}. Both the Vanilla baseline and \sysname{} runs share the same backbone, decoding parameters, and syntax critic budget, so the comparison isolates the effect of \sysname{}'s hint-guided inference pipeline from raw model capability.

\subsection{Main Results}
\label{sec:eval_main}

Table~\ref{tab:main_results} reports the four metrics on the 113 development examples for the three backbones, each in its Vanilla and \sysname{}-equipped configuration.
Across every model and every metric, \sysname{} delivers consistent and substantial gains.

\begin{table*}[t]
  \centering
  \caption{Main results on Spider~2.0--Snow-0212 (113 development examples): Vanilla baseline vs.\ \sysname{} using the GPT-5.5-learned Hint Bank (11 syntax + 37 semantic hints).}
  \label{tab:main_results}
  \begin{tabular}{llcccc}
    \toprule
    Backbone & Configuration & Pass Rate $\uparrow$ & pass@4 $\uparrow$ & Syntax Pass $\uparrow$ & Avg.\ Critics $\downarrow$ \\
    \midrule
    \multirow{2}{*}{Doubao-2.0-lite}
      & Vanilla       & 29.42\% & 46.02\% & 62.17\% & 2.854 \\
      & \sysname{}    & \textbf{49.12\%} & \textbf{64.60\%} & \textbf{96.46\%} & \textbf{0.575} \\
    \midrule
    \multirow{2}{*}{GPT-5}
      & Vanilla       & 42.70\% & 61.06\% & 74.56\% & 2.148 \\
      & \sysname{}    & \textbf{58.85\%} & \textbf{78.76\%} & \textbf{99.12\%} & \textbf{0.602} \\
    \midrule
    \multirow{2}{*}{\textbf{GPT-5.5} (primary)}
      & Vanilla       & 61.95\% & 72.57\% & 96.24\% & 2.790 \\
      & \sysname{}    & \textbf{79.42\%} & \textbf{87.61\%} & \textbf{100.00\%} & \textbf{0.124} \\
    \bottomrule
  \end{tabular}
\end{table*}

For our primary model GPT-5.5, \sysname{} lifts \textbf{pass rate from 61.95\% to 79.42\%} ($+17.47$~pp) and \textbf{pass@4 from 72.57\% to 87.61\%} ($+15.04$~pp), while driving the Snowflake \emph{syntax pass rate} to 100\% and reducing average critic rounds from \textbf{2.790 to 0.124} ($\approx 22{\times}$ reduction). The pass-rate gain is especially noteworthy: it shows that the Hint Bank not only widens the candidate pool (pass@4) but also raises the density of correct queries among all sampled candidates. The near-zero average critic rounds further indicate that essentially all queries are emitted in the correct dialect on the first attempt, so the compiler-feedback loop almost never has to fire during hint-guided inference.

A category-level breakdown of these aggregate gains over the three sub-categories of the development set (\texttt{sf\_other}, \texttt{sf\_local}, \texttt{sf\_bq}) shows that the improvement is concentrated on the harder local-Snowflake and BigQuery-converted subsets while the small \texttt{sf\_other} subset is already near-saturated; full numbers are deferred to Appendix~\ref{app:per_category}.

\subsection{Cross-Model Transferability of the Hint Bank}
\label{sec:eval_transfer}

A central claim of \sysname{} (Section~\ref{sec:methodology}) is that, unlike SFT-style adaptation, a Hint Bank is an \emph{external, model-agnostic} asset: it is learned once with one backbone and can be plugged into other backbones without retraining. We test this directly by reusing the Hint Bank learned by GPT-5.5 on weaker backbones (Doubao-2.0-lite, GPT-5) without any model-specific adaptation. The corresponding rows in Table~\ref{tab:main_results} show that:
\begin{itemize}
    \item On \textbf{Doubao-2.0-lite}, the same bank lifts pass rate from \textbf{29.42\% to 49.12\%} ($+19.70$~pp), pass@4 from \textbf{46.02\% to 64.60\%} ($+18.58$~pp), and the syntax pass rate from \textbf{62.17\% to 96.46\%}, while reducing average critic rounds from \textbf{2.854 to 0.575} ($\approx 5{\times}$).
    \item On \textbf{GPT-5}, pass rate improves from \textbf{42.70\% to 58.85\%} ($+16.15$~pp), pass@4 from \textbf{61.06\% to 78.76\%} ($+17.70$~pp), and the syntax pass rate from \textbf{74.56\% to 99.12\%}, with average critic rounds dropping from \textbf{2.148 to 0.602} ($\approx 3.6{\times}$).
\end{itemize}
The fact that a weaker backbone (Doubao) realizes the largest absolute pass-rate jump indicates that much of the dialect and logic knowledge required by Spider~2.0--Snow can be \emph{externalized} into hints; once externalized, that knowledge transfers across backbones at zero retraining cost---a property that parameter-update-based approaches such as SFT~\cite{scholak2021picard} and DPO~\cite{rafailov2023direct} cannot match.

\subsection{Generalization to Held-Out Examples}
\label{sec:eval_heldout}

The 113 examples used above also served as the \emph{learning} set, so the previous results characterize the Hint Bank's effect on its source distribution. To probe out-of-source generalization we additionally evaluate on the remaining \textbf{434 Spider~2.0--Snow-0212 examples} that the development phase did not learn from; execution correctness is determined by comparing the predicted result set against the official ground-truth result set. To put \sysname{}'s held-out behavior in context, we also report the SQLGenie-style RAG baseline introduced in Section~\ref{sec:experimental_setup}.

\begin{table}[t]
  \centering
  \caption{Held-out evaluation (434 examples not used in development-phase hint learning); all configurations use GPT-5.5 with identical decoding settings. \textbf{SQLGenie-style RAG} is our reproduced SQLGenie-style Example-Bank RAG baseline~\cite{ghosh2025sqlgenie}, injecting top-3 retrieved (question, SQL) pairs.}
  \label{tab:heldout}
  \begin{tabular}{lcccc}
    \toprule
    Configuration & Pass Rate $\uparrow$ & pass@4 $\uparrow$ & Syntax Pass $\uparrow$ & Avg.\ Critics $\downarrow$ \\
    \midrule
    Vanilla     & 56.16\% & 65.90\% & 90.21\% & 1.184 \\
    SQLGenie-style RAG  & 55.30\% & 65.44\% & 85.77\% & 0.446 \\
    \sysname{} & \textbf{58.06\%} & \textbf{67.28\%} & \textbf{99.14\%} & \textbf{0.203} \\
    \bottomrule
  \end{tabular}
\end{table}

Three observations follow. First, the \emph{syntax-level} benefits of \sysname{} transfer almost completely to the held-out split: syntax pass rate climbs from \textbf{90.21\% to 99.14\%} and average compiler-feedback critic rounds drop from \textbf{1.184 to 0.203} ($\approx 5.8{\times}$) once the Hint Bank is enabled. This is intuitive---Snowflake dialect rules captured by Syntax Hints are largely query-independent, so they generalize regardless of whether the example was seen during development-phase learning. Second, the \emph{semantic} gains are noticeably smaller (pass rate $+1.90$~pp, pass@4 $+1.38$~pp) than on the development split. We interpret this as evidence that the held-out and development splits of Spider~2.0--Snow-0212 cover only loosely overlapping semantic distributions: the Semantic Hints distilled in the development phase encode triggers and strategies frequent in those 113 examples that do not all activate on the held-out queries. This suggests that semantic-side benefits scale with how representative the development set is of the target workload---which is exactly the regime in which the Deployment Phase is meant to operate.

Third, the SQLGenie-style RAG baseline is \emph{not} able to convert the same 113-example supervision into held-out gains. It actually trails the Vanilla GPT-5.5 baseline on both execution accuracy and syntax validity ($-0.86$~pp pass rate, $-0.46$~pp pass@4, $-4.44$~pp syntax pass). Although it reduces average critic rounds (from $1.184$ to $0.446$), this efficiency gain does not translate into higher syntax validity or execution accuracy. One possible explanation is that exemplar injection only helps when retrieved examples are sufficiently aligned; otherwise, loosely matched SQL pairs can distract the generator. This is consistent with the fundamental difference between exemplar-injection RAG and structured hint distillation: such a baseline can only paste in retrieved (question, SQL) pairs, so when no nearly-isomorphic example exists in the bank---a common situation under the loose semantic overlap noted above---the retrieved exemplars become a distraction rather than a signal, dragging accuracy below the no-context baseline. \sysname{}, in contrast, distills the same 113 examples into trigger-conditioned, attribution-validated strategies, which generalize beyond verbatim question similarity and mitigate this failure mode.

\subsection{Ablation: Effect of Strategy Attribution}
\label{sec:eval_ablation}

The Hint Management Module's \emph{Strategy Attribution} pass (Section~\ref{sec:management}) is meant to address a specific failure mode: the Hint Bank can be expanded with strategies that are retrieved often but rarely actually useful, conflating ``learned'' with ``useful'' and dragging down inference quality. To isolate its effect, we compare two Hint Banks that differ \emph{only} in whether attribution has been run:
\begin{itemize}
    \item \textbf{w/o Attribution.} The bank produced directly by the Hint Learning Module on the 113 development examples; strategies carry only the learning-time \texttt{recency} signal and \emph{no} $\texttt{eval\_stats}$.
    \item \textbf{with Attribution.} The same bank after a single Strategy Attribution pass populates each strategy's $\texttt{eval\_stats}$ 
    with example-level retrieved/helped/hurt evidence; the inference-time formatting layer then derives success, harm, and inert summaries.
\end{itemize}
Both runs use GPT-5.5 with identical decoding settings; results on the 113 development examples are shown in Table~\ref{tab:ablation_attr}. Both configurations already reach a $100\%$ syntax pass rate, so we omit that column from the table.

\begin{table}[t]
  \centering
  \caption{Effect of Strategy Attribution (GPT-5.5, 113 development examples). Both configurations share the same learned Hint Bank and reach 100\% syntax pass rate (column omitted).}
  \label{tab:ablation_attr}
  \begin{tabular}{lccc}
    \toprule
    Configuration & Pass Rate $\uparrow$ & pass@4 $\uparrow$ & Avg.\ Critics $\downarrow$ \\
    \midrule
    \sysname{} w/o Attribution    & 69.03\% & 79.65\% & 0.334 \\
    \sysname{} with Attribution   & \textbf{79.42\%} & \textbf{87.61\%} & \textbf{0.124} \\
    \bottomrule
  \end{tabular}
\end{table}

Adding the attribution-derived credibility summaries lifts pass rate by \textbf{$+10.39$~pp} and pass@4 by \textbf{$+7.96$~pp}, while further reducing average critic rounds by roughly $2.7{\times}$. Crucially, no new hints are learned between these two configurations---only the planner's view of the bank changes. This confirms our hypothesis from Section~\ref{sec:management}: separating ``learned'' from ``useful'' and giving the planner an explicit per-strategy credibility signal lets it down-weight noisy or rarely-applied strategies and prefer those with empirically strong support, translating directly into higher end-to-end accuracy.

\section{Discussion and Future Work}
\label{sec:discussion}

\subsection{When to Use \sysname{}}
\sysname{} is designed to bridge the gap between static benchmarks and dynamic production environments. It delivers the highest ROI (Return on Investment) in settings where:
\begin{itemize}
  \item \textbf{Environments with Heterogeneous Dialects.} Unlike SFT models that are often overfitted to standard SQL~\cite{scholak2021picard}, \sysname{}'s syntax hints generalize across queries and databases within the same target dialect, while separate dialect-specific syntax registries can be maintained for heterogeneous dialect environments, effectively preventing recurrent compiler failures in strict dialects (e.g., Snowflake) without the need for parameter retraining.
  \item \textbf{Personalized or Domain-Specific Querying.} When user preferences vary significantly or domain-specific knowledge is essential in crafting SQL queries, semantic hints capture these local conventions dynamically. This granular adaptation is often overlooked or too costly to implement via global fine-tuning.
  \item \textbf{Dynamic, Evolving Workloads.} In scenarios requiring continuous adaptation, the Hint Bank evolves naturally as new examples arrive. This avoids the high operational cost of labeling new datasets and retraining models for every schema change.
\end{itemize}
In contrast, for static tasks with limited schema coverage where adaptation is unnecessary, traditional SFT may remain a competitive solution.

\subsection{System Limitations}
We acknowledge several boundaries in our current design:
\begin{itemize}
  \item \textbf{Deployment Phase Not Evaluated.} Our experiments cover only the Development Phase (Phase~I). Because the Deployment Phase (Phase~II) relies on a human-in-the-loop preference signal that public benchmarks such as Spider~2.0--Snow do not provide, we do not run deployment-phase experiments in this paper and leave them as future work.
  \item \textbf{Trigger Sensitivity.} The retrieval of semantic hints relies on the precision of natural-language triggers. Poorly formulated triggers may lead to lower recall or precision in hint selection, impacting the Logic Planning stage.
  \item \textbf{Strategy Sparsity.} While the Strategy Layer prevents conflict loss, in the early stages of deployment (cold start), the system may lack sufficient statistical data to reliably rank competing strategies for rare edge cases.
\end{itemize}

\subsection{Future Directions}
We identify several promising directions to extend this framework:
\begin{itemize}
 \item \textbf{Graph-Based Hint Representation.} Currently, hints are stored as independent entries indexed by natural-language triggers and retrieved at inference time by an LLM-based scope-aware retriever (Section~\ref{sec:inference}). We envision upgrading the Hint Bank to a \textbf{Graph Structure}, where nodes represent business entities (e.g., tables, metrics) and edges represent strategy dependencies. This would enable the system to handle complex, multi-hop reasoning---for instance, realizing that a specific ``Revenue'' calculation strategy depends on a ``Currency Conversion'' rule defined in a separate hint.
  \item \textbf{Multi-Hint Reasoning.} Although the current system uses an LLM-based Logic Planning step to combine retrieved strategies, future work can make multi-hint interaction more explicit and verifiable, for example through graph- or constraint-based reasoning over hint dependencies.
  \item \textbf{Benchmarking against Adaptation Paradigms.} A natural future direction is to conduct systematic comparisons with alternative adaptation methods, including Supervised Fine-Tuning (SFT)~\cite{scholak2021picard}, Direct Preference Optimization (DPO)~\cite{rafailov2023direct}, and Reinforcement Learning (RL)~\cite{pourreza2025reasoning}, to quantify trade-offs in computational cost, adaptability, and interpretability.
  \item \textbf{Deployment and Hint Bank Monitoring.} Deploying \sysname{} in real industry settings and monitoring the Hint Bank's scale is an important next step. Although the Hint Management Module handles merging, deduplication, and conflict resolution, in production the bank can grow with continuous learning; operational monitoring of bank size, retrieval latency, and storage is needed to keep the system sustainable.
  \item \textbf{Knowledge Lifecycle Management.} Real-world deployment requires robust management of the Hint Bank's lifecycle, including mechanisms for pruning outdated hints (Eviction Policies), prioritizing new strategies, and managing storage constraints in long-running systems.
\end{itemize}


\section{Conclusion}

Large language models have made Text-to-SQL increasingly practical, but robust deployment in real database environments remains challenging due to dialect-specific syntax failures, ambiguous business logic, and evolving user preferences. Existing adaptation paradigms each expose a key weakness in this setting: supervised fine-tuning is rigid and costly to refresh, test-time scaling incurs substantial online overhead, and raw retrieval often introduces noisy context rather than actionable guidance.

In this paper, we presented \sysname{}, which reframes Text-to-SQL adaptation as a data management problem over a persistent \emph{Hint Bank}. Instead of repeatedly re-solving the same failures at inference time, \sysname{} converts compiler and execution feedback into reusable \emph{Syntax Hints} and scoped \emph{Semantic Hints}. Its trigger--strategy abstraction further enables the system to preserve and rank competing interpretations explicitly, providing a structured mechanism for handling ambiguity without modifying model parameters.

Experiments on Spider~2.0--Snow-0212 show that this design substantially improves robustness in the development-phase setting. On the 113 supervised development examples of Spider~2.0--Snow-0212, \sysname{} lifts pass rate from 61.95\% to 79.42\% and pass@4 from 72.57\% to 87.61\%, drives the Snowflake syntax pass rate to 100\%, and reduces average compiler-feedback critic rounds from 2.79 to 0.12 per sampled candidate. Crucially, the same Hint Bank, plugged into weaker backbones (Doubao-2.0-lite, GPT-5) without retraining, also produces double-digit pp gains. The held-out evaluation further shows strong syntax-level transfer but more modest semantic gains, suggesting that semantic improvements depend on how well the development examples cover the target workload. These results suggest that explicit, reusable hints can recover much of the benefit of expensive adaptation while retaining low-latency inference and model agnosticism.

More broadly, \sysname{} highlights a systems direction for self-improving database interfaces: keep online query serving lightweight, and move adaptation into the continuous construction and maintenance of external knowledge. While this paper evaluates only the development-phase workflow, the same architecture naturally supports deployment-time batch refresh from accumulated feedback. We hope this perspective helps motivate future work on deployment evaluation, multi-hint reasoning, and long-term knowledge lifecycle management for Text-to-SQL systems.

\section*{Acknowledgments}
This work was supported by the ByteDance ByteBrain team. The authors thank Tieying Zhang and Jianjun Chen for their guidance and support throughout this project.

\bibliographystyle{plainnat}
\bibliography{sample}

\clearpage
\appendix
\section{Atomic \diff{} Schema}
\label{app:hint_diffs}
During the Hint Learning Module's multi-iteration process (Section~\ref{sec:learning_process}), we utilize an LLM to generate atomic \diff{}s between the ground truth SQL and the predicted error traces. This structured schema forces the LLM to isolate individual logic or syntax deviations, facilitating precise hint generation. The prompt enforces the following JSON-like schema, with one object per atomic issue:

\begin{lstlisting}[language={},basicstyle=\ttfamily\small]
DiffID: "<Phrase-or 'GLOBAL'>::<Category>::<running number>"
Phrase: "<verbatim or close paraphrase from Question>"
Category: "<JOIN_TYPE | FILTER_SCOPE | AGG_TIMING |
WINDOW_FRAME | LIMIT_RANK | FORMULA |
COLUMN_CHOICE | UNIT_CAST | SEMANTIC |
GLOBAL_PLAN | OTHER>"
StepRef: "<Gold Step <n> | CTE name | GLOBAL>"
GoldStrategy: "<one concise sentence of Gold's approach>"
GoldSQL: <concise snippet <=200 chars>
WrongStrategy: "<one concise sentence of Pred's approach (mark as unpreferred)>"
WrongSQL: <concise snippet <=200 chars>
Impact: "<one sentence explaining how results or preference deviate>"
\end{lstlisting}

\section{Per-Category Results on the Development Set}
\label{app:per_category}
The 113 development examples of Spider~2.0--Snow-0212 fall into three sub-categories that reflect how each query reaches the Snowflake engine: \texttt{sf\_other} (queries against generic Snowflake-hosted databases, $n{=}6$), \texttt{sf\_local} (queries against locally provisioned Snowflake databases, $n{=}36$), and \texttt{sf\_bq} (queries originally authored on BigQuery and converted to the Snowflake dialect, $n{=}71$). The three subsets pose visibly different challenges: \texttt{sf\_bq} is dialect-heavy because converted queries inherit BigQuery idioms that must be rewritten to Snowflake-compatible forms, while \texttt{sf\_other} contains a small set of relatively standard analytical queries.

Table~\ref{tab:per_category} reports vanilla GPT-5.5 vs.\ \sysname{} (with the GPT-5.5-learned, attribution-finalized Hint Bank) on each sub-category, using \emph{pass@4} and \emph{pass rate}. The aggregate gains in Section~\ref{sec:eval_main} concentrate on the harder sub-categories: \sysname{} delivers \textbf{$+18.31$~pp} pass@4 and \textbf{$+19.72$~pp} pass rate on \texttt{sf\_bq}, and \textbf{$+11.11$~pp} / \textbf{$+15.28$~pp} on \texttt{sf\_local}, while the small \texttt{sf\_other} subset is already near-saturated under the vanilla baseline ($83.33\%$ pass@4) and shows only marginal pass-rate movement. This pattern is consistent with our hypothesis that the Hint Bank captures dialect- and schema-specific knowledge whose marginal value grows with subset difficulty.

\begin{table}[H]
  \centering
  \caption{Per-category results on the 113 development examples (GPT-5.5): Vanilla vs.\ \sysname{}.}
  \label{tab:per_category}
  \begin{tabular}{lcccccc}
    \toprule
    & & \multicolumn{2}{c}{pass@4 $\uparrow$} & \multicolumn{2}{c}{pass rate $\uparrow$} \\
    \cmidrule(lr){3-4} \cmidrule(lr){5-6}
    Category & $n$ & Vanilla & \sysname{} & Vanilla & \sysname{} \\
    \midrule
    \texttt{sf\_other} & 6  & 83.33\% & 83.33\% & 79.17\% & 83.33\% \\
    \texttt{sf\_local} & 36 & 72.22\% & \textbf{83.33\%} & 61.81\% & \textbf{77.08\%} \\
    \texttt{sf\_bq}    & 71 & 71.83\% & \textbf{90.14\%} & 60.56\% & \textbf{80.28\%} \\
    \bottomrule
  \end{tabular}
\end{table}

\end{document}